\documentclass[11pt,a4paper]{article}
\usepackage[margin=24mm]{geometry}
\usepackage[T1]{fontenc}
\usepackage[utf8]{inputenc}
\usepackage{amsmath,amssymb}
\usepackage{mathptmx}
\usepackage{graphicx}
\usepackage{microtype}
\usepackage{setspace}
\usepackage{caption}

\let\HTGMainsection\section
\let\HTGMainsubsection\subsection
\let\HTGMainsubsubsection\subsubsection
\let\HTGMainparagraph\paragraph
\let\HTGMainsubparagraph\subparagraph
\usepackage{titlesec}
\let\HTGSIsection\section
\let\HTGSIsubsection\subsection
\let\HTGSIsubsubsection\subsubsection
\let\HTGSIparagraph\paragraph
\let\HTGSIsubparagraph\subparagraph
\let\section\HTGMainsection
\let\subsection\HTGMainsubsection
\let\subsubsection\HTGMainsubsubsection
\let\paragraph\HTGMainparagraph
\let\subparagraph\HTGMainsubparagraph
\usepackage{booktabs}
\usepackage[percent]{overpic}
\usepackage[super,sort&compress]{natbib}
\usepackage{xcolor}
\usepackage[colorlinks=true,citecolor=black,linkcolor=black,urlcolor=blue]{hyperref}
\usepackage{xcolor}

\hypersetup{pdftitle={Supermoir\'e Formation and Robust Topological Edge State in Helical Trilayer Graphene},pdfauthor={Hong Li and collaborators}}
\graphicspath{{./}{./Figs/}}
\newcommand{\hHTG}{\ensuremath{\mathrm{h}\text{-}\mathrm{HTG}}}
\newcommand{\hbHTG}{\ensuremath{\overline{\mathrm{h}}\text{-}\mathrm{HTG}}}
\newcommand{\AAt}{\ensuremath{\mathrm{AA}_{12}}}
\newcommand{\AAb}{\ensuremath{\mathrm{AA}_{23}}}
\newcommand{\SIsection}[1]{Supplementary Section~\ref{#1}}
\newcommand{\SIfig}[1]{Supplementary Fig.~\ref{#1}}

\newcommand{\MainFig}[1]{Fig.~\ref{#1}}
\newcommand{\MainFigPanel}[2]{Fig.~\ref{#1}#2}
\newcommand{\Id}{\mathbb I}
\newcommand{\nm}{\mathrm{nm}}

\newcommand{\atan}{\operatorname{atan2}}

\newcommand{\moire}{moir\'e}

\begin{document}

{\raggedright\LARGE\bfseries Supermoir\'e Formation and Robust Topological Edge State in Helical Trilayer Graphene\par}
\vspace{8pt}
{\large Hong Li\textsuperscript{1,*}, Huandong Chen\textsuperscript{1,*}, Seongyeon Youn\textsuperscript{2}, Kenji Watanabe\textsuperscript{3}, Takashi Taniguchi\textsuperscript{3}, Trithep Devakul\textsuperscript{2}, Kazuhiro Fujita\textsuperscript{1,$\dagger$} and Abhay Narayan Pasupathy\textsuperscript{1,4,$\ddagger$}\par}
\vspace{5pt}
{\small
\textsuperscript{1}Department of Condensed Matter Physics and Materials Science, Brookhaven National Laboratory, Upton, New York, USA.\\
\textsuperscript{2}Department of Physics, Stanford University, Stanford, California, USA.\\
\textsuperscript{3}National Institute for Materials Science, Tsukuba, Japan.\\
\textsuperscript{4}Department of Physics, Columbia University, New York, New York, USA.\\
\textsuperscript{*}These authors contributed equally.\\
\textsuperscript{$\dagger$}
Corresponding author: kfujita@bnl.gov\\
\textsuperscript{$\ddagger$} Corresponding author: apn2108@columbia.edu

\par}
\vspace{10pt}

{\bfseries
When two materials with nearly matched lattice constants meet, they form commensurate domains separated by incommensurate domain walls that absorb the mismatch. Recently, this picture has been extended to moir\'e lattices: when two moir\'e superlattices have nearly equal periodicities, they can form a ``supermoir\'e'' consisting of commensurate moir\'e domains separated by domain walls. Helical trilayer graphene (HTG) is a particularly interesting example: theory predicts that its commensurate domains are distinct topological insulators, separated by a network of conducting domain walls. Here, we use scanning tunneling microscopy and spectroscopy to visualize supermoir\'e domains and their electronic properties in HTG at a twist angle of $\sim1.7^\circ$. We find atomic-scale evidence of relaxation into supermoir\'e domains, each hosting flat bands separated from remote bands by large gaps of $\sim40$ meV. Domain walls exhibit spectrally uniform edge states along their length, in excellent agreement with calculations for valley-protected topological edge modes. The topological character of the flat bands is intimately tied to wavefunction localization on a graphene sublattice, and we reveal the topological change across a domain boundary by imaging the reversal of sublattice polarization.

\par}
\vspace{6pt}

A defining feature of topological phases is that a gapped bulk can guarantee conducting states at its boundary, whose existence is protected by the topology of the bulk bands. In two-dimensional (2D) topological insulators (TIs), these boundary states are one-dimensional (1D) channels that impose stronger constraints on elastic backscattering than the surface states of three-dimensional TIs \cite{roushan2009topological,BT,BS}. A growing number of 2D systems hosting topological band gaps have been realized \cite{HgTe,Cuizu2013,Yujun2020,Yifan2020,H2020,wu2018quantumspin,serlin2020qah,nuckolls2020chern,Tingxin2021,Cai2023,Tonghang2024,Tang2024,Kang2024,Su2025}. 
However, in practice these edge states are formed at physical sample edges, where strong chemical inhomogeneity and inelastic scattering can degrade transport \cite{Degradeedge}. 
Moir\'e materials also usually feature small intrinsic topological gaps that can limit their observation to very low temperatures. Suitable candidates for high-temperature operation remain scarce \cite{Reis2017,Shumiya2022}. 
Because a topological boundary needs only be an interface where the topological invariant changes, these challenges motivate the search for systems that combine large intrinsic topological band gaps with smooth internal interfaces.

Lattice relaxation in moir\'e systems may offer such a setting. 
At sufficiently small twist angles, lattice relaxation can reconstruct a twisted bilayer into large, commensurate stacking domains separated by narrow domain walls \cite{Jonathan2013,Nam2017,Yoo2019}. 
A similar reconstruction can occur when two moiré lattices are slightly misaligned, producing locally commensurate moiré domains separated by incommensurate domain walls \cite{orderlydisorder,devakul2023helical,guerci2024mosaic,nakatsuji2023multiscale,yang2024multimoire}. 
In this hierarchy, bilayer structures relax at the atomic length scale, while supermoiré (moir\'e of moir\'e) structures relax at the moir\'e length scale. The atomic distortions across the resulting supermoir\'e domain walls vary more gradually in space than those across domain walls in twisted atomic bilayers. 
When neighboring domains have opposite valley Chern numbers and share a bulk energy gap, the domain boundaries connecting them are expected to support counterpropagating, valley-indexed edge modes \cite{Ju2015,huang2018helical,yin2016domainwall,ABCA}. 
Such an intrinsic network of topological edge states with large bandgaps (for moiré materials) has been predicted in helical trilayer graphene (HTG) \cite{devakul2023helical,guerci2024mosaic,nakatsuji2023multiscale}.

Motivated by this idea, recent transport measurements \cite{xia2025topological} have shown evidence for the formation of flat bands at a magic angle of $\sim1.8^\circ$ accompanied by the observation of the anomalous Hall effect, which might arise in a valley-protected topological insulator if the valley symmetry is spontaneously broken. 
Furthermore, scanning single-electron transistor measurements \cite{hoke2026relaxation} have shown evidence for large areas of uniform electronic structure separated by relatively sharp boundaries, in accord with the expected formation of commensurate domains. 
While these and other recent works \cite{xia2026quantized,Hao2026} support the supermoir\'e picture proposed by theory, an atomic-scale validation of this picture, together with its attendant topology, has yet to be achieved. 
Such validation is essential because the energy scales that drive supermoir\'e relaxation are much smaller than those governing commensurate--incommensurate transitions at the atomic scale, so theoretical predictions are correspondingly less reliable. 
In our work we use atomically resolved scanning tunneling microscopy and spectroscopy (STM/S) to achieve this experimentally.

To reveal the HTG structure, we acquire a large-area STM topograph of the surface (Fig.~\ref{fig:one}c). 
Modulations on length scales of $\sim$ 10~nm arise from the graphene moir\'e lattices, whereas triangular domains with side lengths of order 120~nm arise from the interference between the two moir\'e lattices. 
This supermoir\'e network is also clearly visible in the conductance map at 200~meV (Fig.~\ref{fig:one}c, inset). 
The network is relatively uniform in an even larger topograph (\SIsection{sec:large-area}), where a global measurement of the moir\'e period yields an effective average twist angle of $1.74^\circ$ using $a_m=a/(2\sin(\theta/2))$, with $a=0.246$~nm. 
For ideal equal-twist HTG at $1.74^\circ$, the two moir\'e lattices are rotated relative to one another by $1.74^\circ$, producing a regular supermoir\'e with a wavelength of approximately 260~nm, calculated using $a_{mm}=a/[2\sin(\theta/2)]^2$. 
This estimate is approximately twice the observed domain size. 
To understand the smaller size and distorted shape of supermoir\'e in experiment, we first distinguish the upper- and lower-interface moir\'e centers directly in real-space images (Fig.~\ref{fig:one}d). 
Orange dots (\AAt{}) denote AA stacking between the top and middle graphene layers, and blue dots (\AAb{}) denote AA stacking between the middle and bottom layers.

The two sets of moir\'e sites are extracted from the 200~nm-wide region and are shown in Fig.~\ref{fig:one}e, using the method described in \SIsection{sec:site-identification}. 
The results are shown in Fig.~\ref{fig:one}e. 
A supermoir\'e domain is outlined by connecting three AAA sites, where \AAt{} and \AAb{} nearly overlap, as indicated by the green shaded area. 
This coordinate map reveals a systematic mismatch between the upper- and lower-interface moir\'e lattices. 
For example, the strips spanning three rows of twenty AA sites show that the \AAb{} lattice has an approximately 7\% longer real-space period than the \AAt{} lattice. 
Using the spatially averaged primitive lattice vectors extracted from this coordinate map, a rigid overlay of two uniform moir\'e lattices reproduces both the size of the supermoir\'e triangle and the orientations of its three edges relative to the moir\'e lattice vectors (Fig.~\ref{fig:one}f). 
Thus, the supermoir\'e shape reflects the inequivalence and distortion of the two moir\'e lattices.

The measured moir\'e geometry results from, and in turn constrains, the relative deformation of adjacent graphene sheets, but it does not uniquely determine the strain tensor of each sheet. 
Different combinations of layer strain (related by uniform strain applied to all layers) can reproduce similar distortions of the moir\'e and supermoir\'e cells\cite{hoke2026relaxation}. 
Atomic lattice constant measurements in STM are limited by piezo calibrations and drift to a fraction of a percent, and are therefore insensitive to small isotropic strains in the graphene lattice. 
However, because isotropic expansion or contraction of the entire sample has a small effect on the electronic band structure \cite{Huder2018,strainNaoto}, we set it to zero in the analysis that follows. 
The eight components of the four measured moir\'e lattice vectors, together with the measured orientation of a top-layer atomic lattice vector, constrain the six uniaxial-strain parameters and two interface twist angles. The unique solution satisfying these limitations gives twist angles of $1.74^\circ$ for the top--middle interface and $1.61^\circ$ for the middle--bottom interface, with uniaxial strains of approximately $-0.10\%$, $+0.05\%$ and $+0.05\%$ in the top, middle and bottom layers, respectively (\SIsection{sec:strain_config}). Therefore, the reduced size and distortion of the supermoir\'e arise from a combination of an approximately $0.12^\circ$ interface twist-angle mismatch and heterostrain. 
These small strain values are expected to marginally increase the flat-band width without affecting the topology of the band structure\cite{strainFuliang,strainNaoto}. 

Even at a reduced size of approximately 120~nm, the supermoir\'e domains remain large enough for substantial relaxation. 
This is already evident in Fig.~\ref{fig:one}e,f: within a single domain, the experimental moir\'e lattice is much more uniform than in the rigid overlay, despite their similar supermoir\'e cell shapes. 
Calculations also predict local commensuration for consecutive twists, even when a small twist-angle mismatch reduces the supermoir\'e scale\cite{Zhu2020,nakatsuji2023multiscale,yang2024multimoire}. 
To quantify the relaxation, we directly extract the real-space lateral shift between the upper- and lower-interface AA sites, defined as $\mathbf d(\mathbf r)=\mathbf r(\mathrm{AA}_{23})-\mathbf r(\mathrm{AA}_{12})$ (Fig.~\ref{fig:one}d). 
We reduce each shift modulo a shared reference moir\'e lattice into its unit cell. 
Theory predicts that $\mathbf d(\mathbf r)$ remains close to the preferred values $\boldsymbol\delta_{\pm}=\pm(\mathbf a_1+\mathbf a_2)/3$ (mod $\Lambda$), located at the centroids of the two triangular halves of the unit cell. Here, $\mathbf a_1$ and $\mathbf a_2$ are primitive real-space moir\'e lattice vectors and $\Lambda=\{n_1\mathbf a_1+n_2\mathbf a_2:n_1,n_2\in\mathbb Z\}$. 
Using the coordinates in Fig.~\ref{fig:one}e, we map $\mathbf d$ in real space (Fig.~\ref{fig:one}g), with colors approaching cyan near $\boldsymbol\delta_+$ and pink near $\boldsymbol\delta_-$. 
As expected, $\mathbf d(\mathbf r)$ clusters near these preferred values within the domains, producing distinct colors for neighboring supermoir\'e domains. 
This clustering is also evident in the statistical distribution of $\mathbf d$ in a unit cell (Fig.~\ref{fig:one}h), where the vector endpoints concentrate near the center of moir\'e unit cell triangles. 
The two preferred values of $\mathbf{d}$ correspond to the \hHTG{} and \hbHTG{} domains predicted by theory\cite{devakul2023helical,guerci2024mosaic}. 
In particular, the $\mathbf{d}$ vectors within the domain interior (at least two moir\'e unit cells away from the domain boundaries) are confined to a region covering only $\sim$9\% of the unit cell area. 
In the rigid-rotation case, the $\mathbf{d}$ vectors sample the entire unit cell uniformly. 
Their concentration, rather than a continuous distribution of rigid-lattice registries, directly supports reconstruction into locally uniform domains. 

A well-defined Bloch band structure exists within a single uniform supermoir\'e domain. 
For instance, we simulate the band structure of a uniformly relaxed HTG domain with $\sim$1.7$^\circ$ equal-twist stacking (Fig.~\ref{fig:two}a, \SIsection{sec:theory_singledomain}), in which the two flat bands are isolated from the remote bands and each carries a nonzero valley Chern number. 
This band structure is reflected in density-of-states (DOS) spectra, which can be probed using the differential-conductance ($\mathrm dI/\mathrm dV$) spectroscopy. 
To examine the DOS spectra in our sample, we locate an area within a single supermoir\'e domain (Fig.~\ref{fig:two}b). 
Three inequivalent sites exhibit distinct topographic heights, forming a hexagonal structure similar to the relaxed model shown in Fig.~\ref{fig:one}b. 
They are  assigned as AAB, BAA and ABA as described below. 
The stacking letters describe relative layer registries and are ordered from the exposed top layer (1) to the bottom layer (3), as shown in Fig.~\ref{fig:one}a.

The assignments of the three stacking sites are established spectroscopically. 
The measured d$I$/d$V$ spectra at zero gate voltage (Fig.~\ref{fig:two}c), which are proportional to the local density of states (LDOS), reproduce the main stacking-dependent features of the calculated surface LDOS at neutral filling (Fig.~\ref{fig:two}d). 
The approximately 10~meV energy shift between experiment and theory arises from unintentional doping, which is commonly observed in two-dimensional heterostructures\cite{Xie2019,Tilak2021}. 
The two-peak structure near the Fermi level is associated with the two narrow flat bands in Fig.~\ref{fig:two}a, while peaks near $\pm130$~meV arise from remote bands. 
Near the flat-band energy, AAB carries the largest spectral weight, followed by ABA and BAA. 
This is consistent with the intuition of enhanced surface LDOS near the upper-interface AA regions, similar to the AA-centered flat-band LDOS in TBG\cite{Kerelsky2019}. 
Furthermore, the experimental and calculated real-space maps support the same assignment: the flat-band signal is the strongest at AAB and weakest at BAA (Fig.~\ref{fig:two}e).

The experimental remote-band gaps appear as conductance minima rather than hard zeros when they lie away from the Fermi level. 
This behavior arises from finite excitation lifetimes and inelastic tunneling far from the Fermi level with possible contributions of incoherent states. However, when gating places the electron-side remote-band gap at the Fermi level, the domain-interior conductance approaches the experimental noise floor (\SIsection{sec:spectroscopy}). 
In addition, we performed gate-dependent differential conductance spectroscopy, where no new correlation-induced spectral features are resolved between full hole filling ($\nu=-4$) and full electron filling ($\nu=+4$) (\SIfig{fig:gate-spectroscopy}). 
This is possibly because of the deviation of the twist angles from the magic angle condition at which the bandwidth is theoretically minimized.

The opposite $\mathbf{d}$ vectors in neighboring supermoir\'e domains should give rise to opposite valley Chern numbers for their Bloch bands and, consequently, topological edge states at the supermoir\'e domain wall. 
However, structural distortions at the domain wall itself may also produce prominent dispersive states of non-topological origin\cite{Ouyang2026,Nichele_2016,Oles2018}. 
To assess the origin of any domain wall states, we calculate the band structure fitted to the measured moir\'e sites (Fig.~\ref{fig:three}a, \SIsection{sec:theory_domainwall}), where translational periodicity along the wall defines a 1D momentum. 
The calculated spectrum contains dense projected domain bands and dispersive domain-wall branches that connect the flat and remote bands across the isolation gaps (Fig.~\ref{fig:three}b), consistent with theoretical predictions for reconstructed HTG\cite{nakatsuji2023multiscale,guerci2024mosaic}. 
The calculated LDOS is correspondingly enhanced at the wall within both remote-band gaps, owing to the topological dispersive 1D edge states (Fig.~\ref{fig:three}c). 
Note that the edge states within the band gap have low LDOS amplitude ($\sim$3\% of the flat-band LDOS) and exhibit a flat spectral profile rather than a peak. 
These features arise from the presence of a single edge state mode whose band velocity is much higher than that of the bulk bands, resulting in a much smaller LDOS $\propto 1/\mathbf{v}_\text{F}$.

Significantly, we observe the same phenomenon experimentally. 
By applying a gate voltage, we bring the hole-side and electron-side isolation gaps to the Fermi level at $\nu=-4$ and $\nu=+4$, respectively. 
At both fillings, spectra in the domain interior show strongly suppressed conductance over an interval spanning tens of meV, whereas spectra at the wall exhibit greater in-gap weight (Fig.~\ref{fig:three}d,e). 
This in-gap LDOS is small, $\sim$5\% of the flat-band LDOS, and exhibits a relatively flat spectral profile. 
Both features are consistent with a single dispersive edge mode, as predicted by theory. 
The comparison also shows reduced remote-band peak intensity and a redistribution of spectral weight away from the gap, in remarkable agreement with the calculated domain-wall LDOS near the gap. 
A differential conductance map near the Fermi level further reveals enhanced weight along the domain walls (Fig.~\ref{fig:three}f). 
The wall response retains moir\'e-scale modulation rather than forming a featureless line.

A calculated in-gap LDOS map reproduces this modulation, with wall-associated weight extending across approximately three to four moir\'e periods (Fig.~\ref{fig:three}g), matching the length scale of the moir\'e distortion fitted to the measured moir\'e sites (Fig.~\ref{fig:three}a). 
Faint dot-like spectral weight is visible inside the experimental domains, but its magnitude is small compared with the band-edge signal. 
In particular, the low-energy domain spectrum approaches the noise floor (Fig.~\ref{fig:three}e, inset). 
The line cut across the wall further shows how the low-conductance interval acquires additional weight between the two domain interiors (Fig.~\ref{fig:three}h). 
In addition, the edge-state LDOS amplitudes and widths differ among the three domain walls. 
The spatial profile of the edge state spectral intensity depends on both the domain wall profile and the intrinsic penetration length into the gapped bulk. 
The latter scales as $\hbar v_{\perp}/\Delta$, where $v_{\perp}$ is the relevant bulk velocity normal to the wall and $\Delta$ is the bulk gap energy scale. 
Different domain wall widths can therefore lead to different spatial extents of the edge states. 
The domain wall width depends on its orientation relative to the underlying lattice, analogous to the orientation-dependent widths of solitons in small-angle twisted bilayer graphene \cite{Jonathan2013}. 
The domain wall shown in Fig.~\ref{fig:three}a, which is aligned with the moir\'e lattice, appears sharpest in the STM topography and exhibits a narrower spatial profile of the edge state spectral intensity. 
Given the relatively small differences in atomic distortion among the three edges and the much larger length scale of the domain walls, the observed variation in confinement suggests that the edge states respond primarily to the moir\'e-scale edge profile. 
This behavior is consistent with a topological edge-state interpretation rather than trivial electronic states caused by atomic distortion.

The agreement between the observed in-gap states at supermoir\'e domain walls and the calculations supports a topological origin. 
We next provide further evidence for this interpretation. 
In the relaxed model, \hbHTG{} and \hHTG{} are $C_{2z}$-related structures. 
The breaking of $C_{2z}$ within each domain permits a nonzero valley-resolved Chern number and nonzero Berry curvature\cite{devakul2023helical,xia2025topological}. 
The sublattice-selective electronic response provides an experimentally accessible signature of this asymmetry. 
This motivates a comparison of carbon-site LDOS across the supermoir\'e domain wall identified above.

\begin{figure}[p]
\centering
\includegraphics[width=\textwidth]{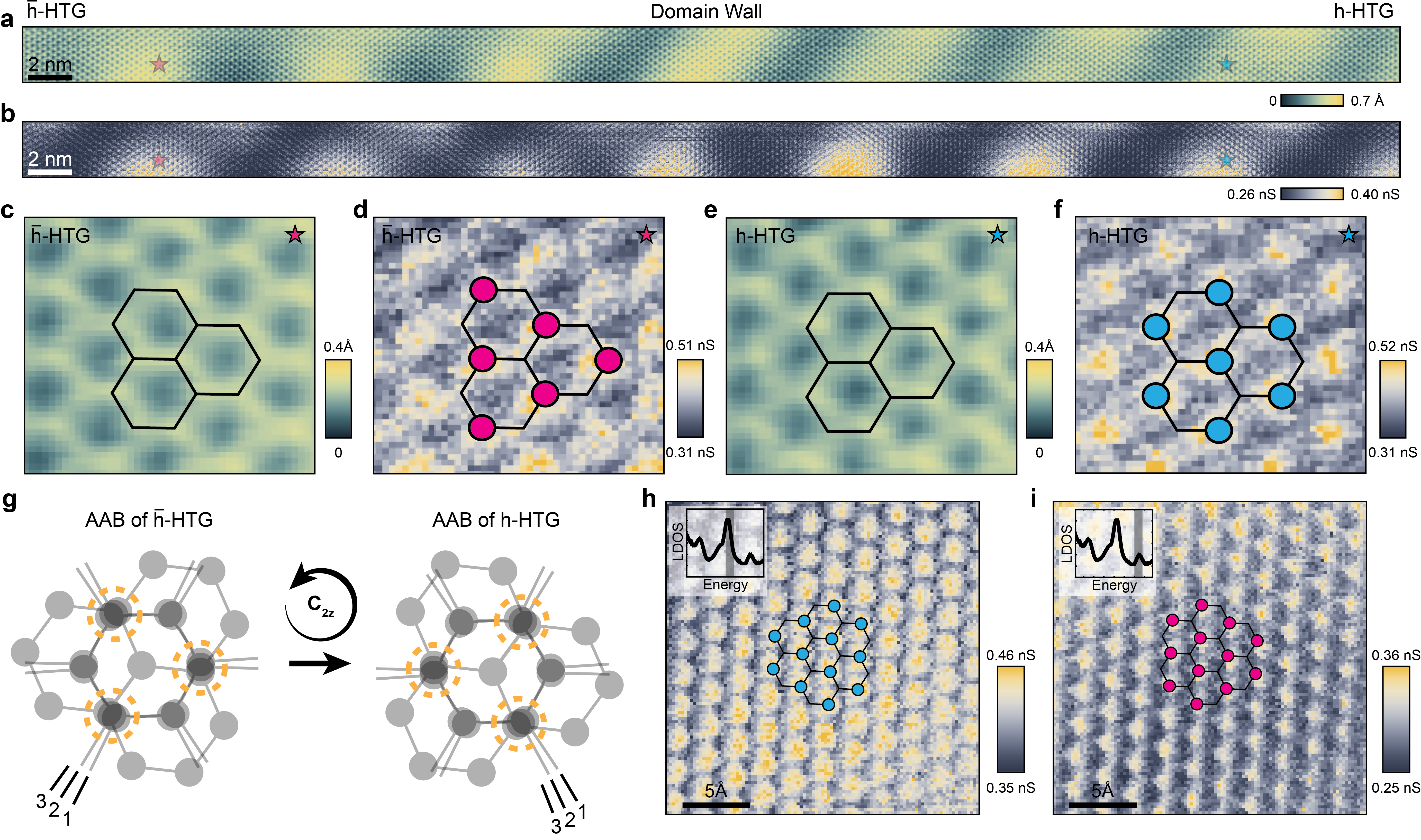}
\caption{\textbf{Atomic sublattice polarization across domains and between bands.}
\textbf{a,b}, Simultaneously acquired atomically resolved topograph at a setpoint bias of 1~V (\textbf{a}) and dI/dV map at a flat-band energy (0 meV) (\textbf{b}) across an \hbHTG{}--\hHTG{} wall. Magenta and cyan stars identify the AAB regions enlarged in \textbf{c--f}.
\textbf{c,d}, Enlarged topograph (\textbf{c}) and conductance map (\textbf{d}) in \hbHTG{}. A common honeycomb reference registers the top-layer carbon lattice; magenta markers identify the sublattice with enhanced LDOS.
\textbf{e,f}, Corresponding topograph (\textbf{e}) and dI/dV map (\textbf{f}) in \hHTG{}, where cyan markers identify the complementary sublattice.
\textbf{g}, Atomic stacking schematics of the corresponding AAB registries in the two domain types, related by $C_{2z}$. Layer numbers are shown, and orange dashed circles highlight the locally aligned columns whose relation to the exposed sublattice changes under the rotation.
\textbf{h,i}, Registered conductance maps of the same AAB region within one domain at the flat-band energy of 20~meV (\textbf{h}) and the positive remote-band energy of 120~meV (\textbf{i}). The honeycomb overlays show a reversal of the preferred sublattice, highlighted in cyan and magenta, respectively. The insets show a local spectrum with the sampled energy regions shaded. STM acquisition settings: \textbf{a--f}, sample bias, 1~V; current setpoint, 500~pA; excitation voltage, 10~mV; back-gate voltage, 0~V. \textbf{h,i}, sample bias, 200~mV; current setpoint, 50~pA; excitation voltage, 10~mV; back-gate voltage, 0~V.}
\label{fig:four}
\end{figure}

To examine this experimentally, we simultaneously acquire an atomically resolved topograph at a setpoint bias of 1~V and a differential conductance map at a flat-band energy across a domain wall (Fig.~\ref{fig:four}a,b). 
The topograph at 1~V supplies a structure-sensitive reference for the exposed carbon lattice, enabling a comparison of the sublattices at two AAB sites on opposite sides of the domain wall. 
The differential conductance map provides LDOS contrast between the two graphene sublattices, revealing potential sublattice inequivalence. 
The enlarged AAB regions in the \hbHTG{} and \hHTG{} domains exhibit similar honeycomb topographic contrast (Fig.~\ref{fig:four}c,e). 
Referenced to the same registered lattice, their flat-band differential conductance maps select complementary carbon sites: the dominant signal lies on the A sublattice in \hbHTG{} and on the B sublattice in \hHTG{} (Fig.~\ref{fig:four}d,f). 
We label these A and B using a common top-layer lattice convention. The observed reversal is consistent with the change in local interlayer registry and the $C_{2z}$ relation between the domains (Fig.~\ref{fig:four}g). 
Together with the opposite lateral shift vectors in Fig.~\ref{fig:one}g,h, it links the moir\'e-scale reconstruction to atomic-scale electronic $C_{2z}$ symmetry breaking.

As a further check, we simultaneously map the AAB site within one domain at a flat-band energy and a positive remote-band energy (Fig.~\ref{fig:four}h,i). 
The preferred sublattice reverses between the two energies while the honeycomb reference remains fixed. 
This energy dependence supports an electronic wavefunction origin for the contrast.

Our measurements connect three length scales in the HTG: the carbon lattice, the local moir\'e unit cell and the reconstructed supermoir\'e network. 
The lateral shifts cluster into two $C_{2z}$-related cases and remain nearly uniform inside individual supermoir\'e domains. 
When gating places a remote-band gap at the Fermi level, additional spectral weight follows the boundaries between these domains, in good agreement with the calculated interface spectrum and spatial LDOS. 
Complementary atomic sublattice polarization further links the domain registries to their electronic $C_{2z}$ symmetry breaking. 
Together, these observations provide microscopic evidence for the predicted valley-topological domain-wall network. 
In addition, in-gap states are observed where the continuous HTG sheet crosses an hBN substrate step (\SIfig{fig:step_edge}). 
Their stronger response highlights the sensitivity of interface spectra to local confinement and electrostatics; the topological character of these step-associated states remains open. 
Future measurements combining local spectroscopy with transport can test how this reconstructed network carries current and whether the large intrinsic isolation gaps can support robust valley transport at elevated temperatures.

\section*{Methods}

\subsection*{Device fabrication}
The helical twisted trilayer graphene device was prepared using a procedure slightly modified from our previous report\cite{chen2026guiding}. A freshly cleaved hBN flake and three monolayer graphene flakes cut by atomic force microscopy (AFM) were sequentially picked up from SiO$_2$ substrates using a low-molecular-weight polycarbonate (PC) stamp at $80\,^{\circ}\mathrm{C}$, following a standard cut-and-stack method. The crystal axes of hBN and graphene were intentionally misaligned to avoid an hBN/graphene moir\'e pattern. The device was temporarily released onto a clean SiO$_2$ substrate by melting the PC film at $140\,^{\circ}\mathrm{C}$. After the PC film was dissolved in chloroform, the device stack was picked up again using a polyvinyl chloride (PVC) stamp at $80\,^{\circ}\mathrm{C}$, with a low release speed to minimize process-induced strain. It was then released onto a planar polydimethylsiloxane (PDMS) stamp for flipping at $120\,^{\circ}\mathrm{C}$.

The flipped device was aligned and transferred onto the central region of a SiO$_2$ substrate encoded with a guiding pattern at $80\,^{\circ}\mathrm{C}$ to facilitate STM navigation to the two-dimensional device\cite{chen2026guiding}. Process temperatures were kept at or below $140\,^{\circ}\mathrm{C}$ to minimize undesired relaxation. The surface of graphene device was thoroughly cleaned by N-methyl-2-pyrrolidone (NMP)-based organic solvents and an AFM tip in a contact mode (FX40, Park Systems), followed by metal deposition of Ti/Au = 5/80 by Ebeam evaporation through a thin silicon shadow mask. The shadow mask was fabricated from a $100~\mu\mathrm{m}$-thick silicon wafer by cryogenic silicon deep etching using a 100~nm-thick Ni film as the etch mask.

\subsection*{STM and spectroscopy}
Measurements were performed on the HTG device at 4.5~K using a commercial Unisoku-1300 ultrahigh-vacuum STM. A chemically etched tungsten tip was cleaned and calibrated on an electron-beam-deposited gold surface before measurement. Differential conductance is used as a proxy for the surface LDOS under the usual assumptions about the tunneling matrix element and tip density of states\cite{tersoff1985stm}. The filling $\nu$ is measured relative to charge neutrality in carriers per moir\'e cell; full electron and hole fillings of the respective four-flavour bands correspond to $\nu=+4$ and $\nu=-4$.

\section*{Acknowledgements}
Work at Brookhaven National Laboratory (BNL) is supported by the Office of Basic Energy Sciences, Materials Sciences and Engineering Division, U.S. Department of Energy (DOE) under contract No. DE-SC0012704.
This research used Nanofabrication facilities of the Center for Functional Nanomaterials (CFN), which is a U.S. DOE Office of Science Facility, at BNL under contract No. DE-SC0012704. TD and SY at Stanford university acknowledge support from the Air Force Office of Scientific Research under award
number FA9550-25-1-0343 and the Sloan Research Foundation.

\clearpage

\begin{figure}[p]
\centering
\includegraphics[width=\textwidth]{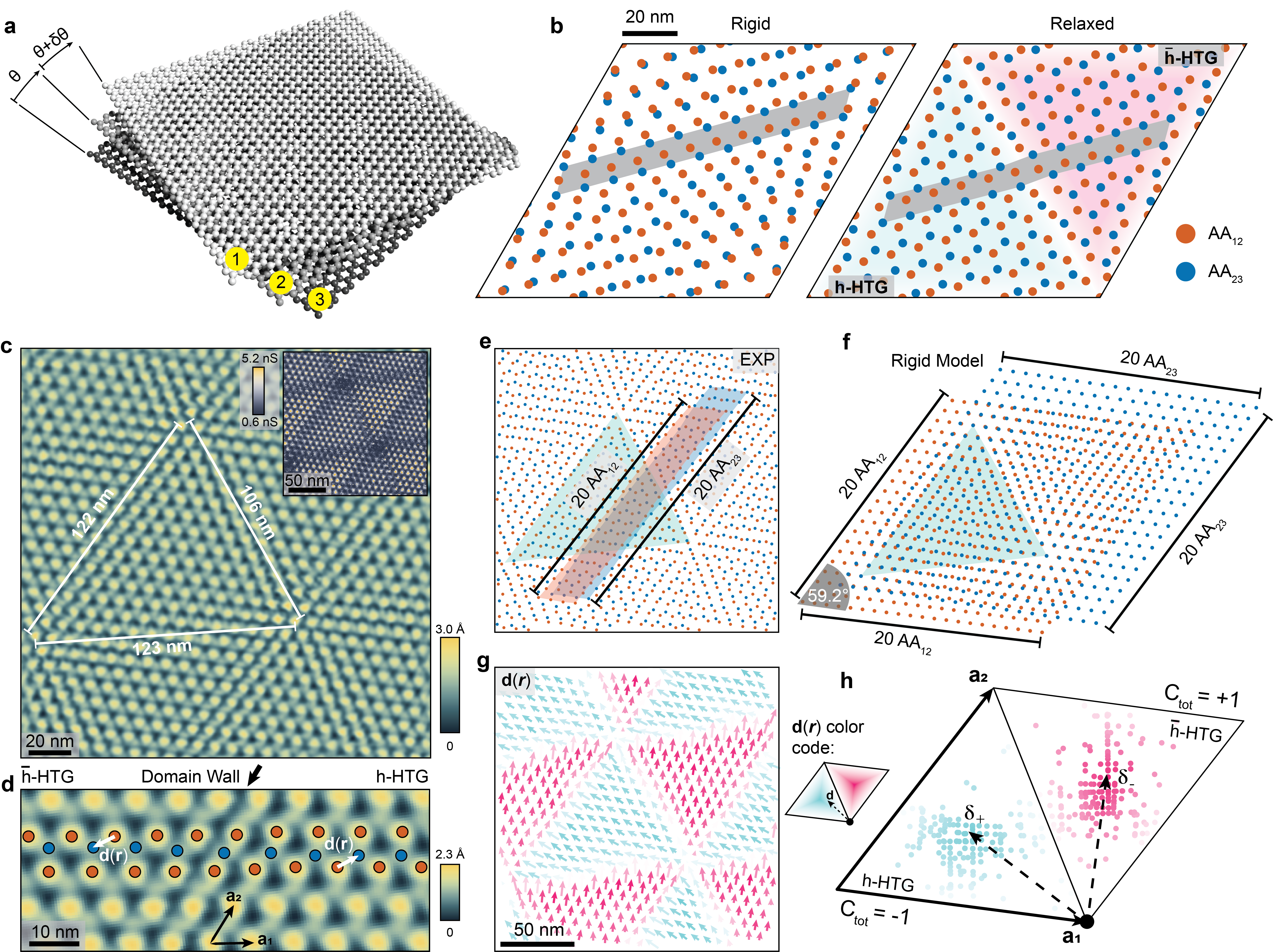}
\caption{\textbf{Structure and relaxation of helical trilayer graphene.}
\textbf{a}, Ball representation of three graphene sheets, numbered from the exposed top layer (1) to the bottom layer (3), with consecutive rotations $\theta$ and $\theta+\delta\theta$ in the same direction.
\textbf{b}, Rigid and relaxed arrangements of upper-interface \AAt{} centers (orange) and lower-interface \AAb{} centers (blue). Relaxation favors \hHTG{} (cyan) and \hbHTG{} (pink) domains with opposite lateral shifts; grey strips mark representative wall-crossing regions, illustrating the distortion in the relaxed case. \textbf{c}, A 200~nm-wide STM topograph filtered at the moir\'e wave vectors, showing triangular supermoir\'e domains with a representative side length of 120~nm. The inset shows a dI/dV map at 200~meV in which the domain network is more clearly visible.
\textbf{d}, Enlarged topograph across an \hbHTG{}--\hHTG{} wall. Orange and blue dots mark \AAt{} and \AAb{} centers, respectively, while white arrows illustrate $\mathbf d(\mathbf r)$, directed from \AAt{} to \AAb{} and reduced modulo the reference moir\'e lattice. An example pair of local moir\'e primitive vectors $\mathbf{a}_1$ and $\mathbf{a}_2$ is indicated by black arrows.
\textbf{e}, Extracted experimental center positions, with orange and blue dots denoting \AAt{} and \AAb{}, respectively. The shaded triangle outlines a supermoir\'e domain, and the elongated strips compare lengths spanning twenty sites of the two moir\'e lattices.
\textbf{f}, Rigid-lattice overlay incorporating the approximately 7\% moir\'e wavelength mismatch and the indicated $59.2^\circ$ angle between the upper-interface primitive real-space moir\'e lattice vectors. The shaded triangle illustrates the resulting beat-pattern scale. \textbf{g}, Experimentally measured $\mathbf d(\mathbf r)$ field in real space. The direction and length of each vector are color-coded according to its endpoint in the moir\'e cell, as shown in the adjacent diamond. Pink and cyan distinguish the preferred registries, with intermediate colors approaching white.
\textbf{h}, Reduced $\mathbf{d}(\mathbf{r})$ (dots) in the shared moir\'e unit cell. Clusters occur near the two preferred registry cases and their lattice-equivalent positions. The dots use the same color code as the corresponding vectors in \textbf{g}. The $C_{\mathrm{tot}}=\pm1$ annotations are model assignments for the sum of the two central-band Chern numbers for one chosen valley and spin, rather than measured invariants or a charge Chern number summed over valleys. Primitive vectors $\mathbf a_1$ and $\mathbf a_2$ and domain labels are indicated. STM acquisition settings: \textbf{c,d}, sample bias, 200~mV; current setpoint, 200~pA; back-gate voltage, +100~V.}
\label{fig:one}
\end{figure}
\clearpage

\begin{figure}[p]
\centering
\includegraphics[width=\textwidth]{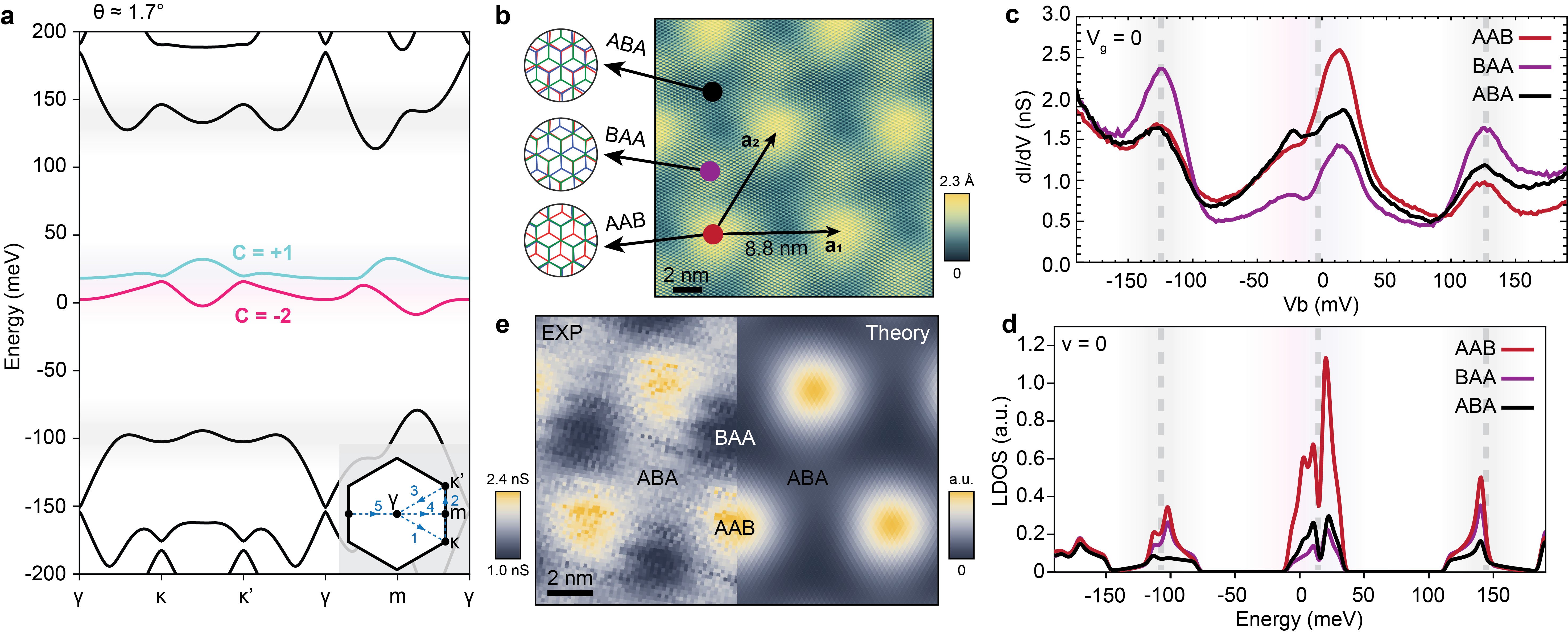}
\caption{\textbf{Band structure and spectra of helical trilayer graphene.} \textbf{a}, Calculated bands for a single relaxed domain and the $K$ valley at $\theta=1.74^\circ$, along the moir\'e Brillouin-zone path illustrated in the inset. The two central bands have model Chern numbers $+1$ (cyan) and $-2$ (pink) for one spin; black curves denote remote bands. These Chern numbers do not represent a total charge Chern number summed over both valleys. 
\textbf{b}, Atomically resolved STM topograph with an 8.8~nm local moir\'e period. Markers identify AAB (red), BAA (purple) and ABA (black). The adjacent schematics illustrate the relative layer registries. Arrows indicate the local primitive vectors $\mathbf a_1$ and $\mathbf a_2$.
\textbf{c}, Measured $\mathrm dI/\mathrm dV$ spectra at the three sites for $V_{\mathrm g}=0$, using the same site colors.
\textbf{d}, Calculated surface LDOS spectra at neutral filling, $\nu=0$, for the corresponding sites. Dashed lines in \textbf{c,d} mark representative flat- and remote-band peaks, allowing for the energy offset between experiment and calculation.
\textbf{e}, Experimental dI/dV map (left) and calculated LDOS map (right) at 0~meV, showing the strongest surface weight at AAB. STM acquisition settings: \textbf{b}, sample bias, 200~mV; current setpoint, 50~pA; back-gate voltage, +100~V. \textbf{c,e}, sample bias, 200~mV, current setpoint, 200~pA, excitation voltage, 2~mV, back-gate voltage, 0~V. The calculated LDOS in \textbf{d,e} uses Gaussian broadening with a standard deviation of 2~meV. 
}
\label{fig:two}
\end{figure}

\clearpage

\begin{figure}[p]
\centering
\includegraphics[width=\textwidth]{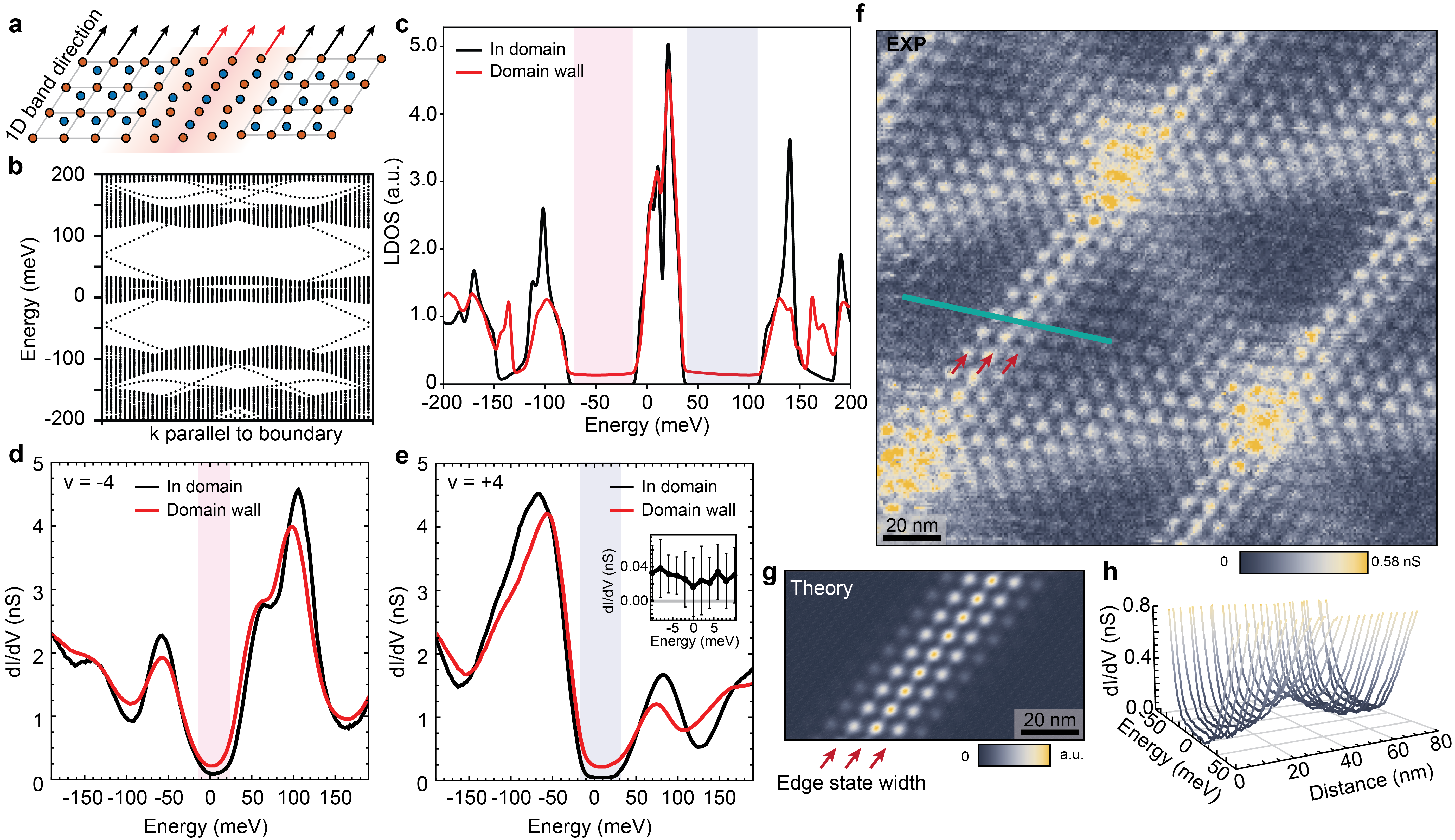}
\caption{\textbf{In-gap states at supermoir\'e domain walls.}
\textbf{a}, Domain-wall geometry used for the calculation. Orange and blue centers represent the two moir\'e lattices; arrows indicate the direction of translational periodicity and one-dimensional momentum parallel to the boundary. Red arrows and shading identify the wall region.
\textbf{b}, Calculated energy spectrum versus momentum parallel to the wall.
\textbf{c}, Calculated LDOS in a domain (black) and at the wall (red). Pink and blue shading mark the hole-side and electron-side isolation gaps, respectively.
\textbf{d,e}, Measured dI/dV spectra in a domain (black) and at a wall (red), with the Fermi level in the hole-side gap at $\nu=-4$ (\textbf{d}) and the electron-side gap at $\nu=+4$ (\textbf{e}). The shaded windows highlight the corresponding low-LDOS intervals. The inset in \textbf{e} shows an enlarged low-energy domain-interior spectrum with error bars representing standard deviations calculated from the spectra used in the spatial average.
\textbf{f}, Experimental in-gap dI/dV map at $E=0$~meV and $\nu=+4$ showing the boundary network and junctions. The green line indicates the path for \textbf{h}. Red arrows indicate the transverse region of enhanced weight.
\textbf{g}, Calculated spatial LDOS inside the electron-side remote-band gap at $E=50$~meV, referenced to the $\nu=0$ spectrum in \textbf{c}. The arrows indicate the transverse extent of the wall-associated weight, spanning approximately three to four moir\'e periods.
\textbf{h}, Spectroscopic line cut along the green path in \textbf{f}, showing the evolution of the gap from one domain interior through the wall to the next. Trace positions indicate distance, while trace height and color indicate differential conductance over the displayed range. STM acquisition settings: \textbf{d}, sample bias, 200~mV; current setpoint, 200~pA; back-gate voltage, -100~V ($\nu=-4$); excitation voltage, 2 mV. \textbf{e,f,h}, sample bias, 200~mV; current setpoint, 200~pA; back-gate voltage, $+100$~V ($\nu=+4$); excitation voltage, 2 mV. The calculated LDOS in \textbf{c,g} uses Gaussian broadening with a standard deviation of 2~meV.}
\label{fig:three}
\end{figure}
\clearpage

\clearpage

\clearpage
\let\section\HTGSIsection
\let\subsection\HTGSIsubsection
\let\subsubsection\HTGSIsubsubsection
\let\paragraph\HTGSIparagraph
\let\subparagraph\HTGSIsubparagraph
\titlespacing*{\section}{0pt}{12pt}{5pt}
\setcounter{page}{1}
\setcounter{figure}{0}
\setcounter{table}{0}
\setcounter{equation}{0}
\setcounter{section}{0}
\setcounter{subsection}{0}
\setcounter{subsubsection}{0}
\renewcommand{\thesection}{S\arabic{section}}
\renewcommand{\thesubsection}{\thesection.\arabic{subsection}}
\renewcommand{\thefigure}{S\arabic{figure}}
\renewcommand{\thetable}{S\arabic{table}}
\renewcommand{\theequation}{S\arabic{equation}}
\renewcommand{\thepage}{S\arabic{page}}
\renewcommand{\figurename}{Supplementary Figure}
\renewcommand{\tablename}{Supplementary Table}
\renewcommand{\theHsection}{S\arabic{section}}
\renewcommand{\theHsubsection}{\theHsection.\arabic{subsection}}
\renewcommand{\theHsubsubsection}{\theHsubsection.\arabic{subsubsection}}
\renewcommand{\theHfigure}{S\arabic{figure}}
\renewcommand{\theHtable}{S\arabic{table}}
\renewcommand{\theHequation}{S\arabic{equation}}

\setcounter{figure}{0}
\renewcommand{\thefigure}{S\arabic{figure}}
\setcounter{section}{0}
\renewcommand{\thesection}{S\arabic{section}}

{\raggedright\large\bfseries Supplementary Information\par}
\vspace{6pt}
{\raggedright\LARGE\bfseries Supermoir\'e Formation and Robust Topological Edge State in Helical Trilayer Graphene\par}
\vspace{8pt}
{\large Hong Li\textsuperscript{1,*}, Huandong Chen\textsuperscript{1,*}, Seongyeon Youn\textsuperscript{2}, Kenji Watanabe\textsuperscript{3}, Takashi Taniguchi\textsuperscript{3}, Trithep Devakul\textsuperscript{2}, Kazuhiro Fujita\textsuperscript{1,$\dagger$} and Abhay Narayan Pasupathy\textsuperscript{1,4,$\ddagger$}\par}
\vspace{5pt}
{\small
\textsuperscript{1}Department of Condensed Matter Physics and Materials Science, Brookhaven National Laboratory, Upton, New York, USA.\\
\textsuperscript{2}Department of Physics, Stanford University, Stanford, California, USA.\\
\textsuperscript{3}National Institute for Materials Science, Tsukuba, Japan.\\
\textsuperscript{4}Department of Physics, Columbia University, New York, New York, USA.\\
\textsuperscript{*}These authors contributed equally.\\
\textsuperscript{$\dagger$}
Corresponding author: kfujita@bnl.gov\\
\textsuperscript{$\ddagger$} Corresponding author: apn2108@columbia.edu
\par}
\vspace{10pt}

\section{Theory}
\subsection{Band-structure calculations within a single domain}
\label{sec:theory_singledomain}
In this section, we consider equally twisted helical trilayer graphene with layer angles $(\theta_1,\theta_2,\theta_3)=(-\theta,0,\theta)$, where layers 1, 2, and 3 correspond to the top, middle, and bottom graphene layers, respectively. In helical trilayer graphene, lattice relaxation locally modifies the twist angles and lattice constants, allowing the two  moir\'e patterns to be commensurate and restoring moir\'e periodicity within each supermoir\'e domain~\cite{SIbib-devakul2023helical}. To model the locally commensurate geometry of a relaxed
supermoir\'e domain, we introduce the lattice scaling
factors
$(\lambda_1,\lambda_2,\lambda_3)=(1,\frac1{\cos\theta},1)$,
which make the two interlayer moir\'e reciprocal
lattices coincide. The resulting real- and reciprocal-lattice vectors of layer $l$, expressed as row vectors, are
\begin{equation}
 A_l = \lambda_l A_0 R(\theta_l)^T,\quad G_l = 2\pi(A_l^{-1})^T,\quad A_0 = a_0
 \begin{pmatrix}
 1&0\\
 1/2&\sqrt{3}/2
 \end{pmatrix},
\end{equation}
where $R(\theta)$ denotes a counterclockwise rotation and $a_0=2.46\text{\AA}$ is the graphene lattice constant. Throughout our calculations, we set $\theta=1.74^\circ$. 

We model the supermoir\'e domain bulk using a moir\'e-periodic effective continuum Hamiltonian for valley $K$~\cite{SIbib-BM,SIbib-devakul2023helical,SIbib-xia2025topological,SIbib-Yves2025},
\begin{equation}
    {H}_K=
    \begin{pmatrix}
        h_1 & T_{12}^{\dagger}(\mathbf{r}-\mathbf{d}_{12}) & 0\\
        T_{12}(\mathbf{r}-\mathbf{d}_{12}) & h_2 & T_{23}^{\dagger}(\mathbf{r}-\mathbf{d}_{23})\\
        0 & T_{23}(\mathbf{r}-\mathbf{d}_{23}) & h_3
    \end{pmatrix}
    \label{eq:HTG-Hamiltonian}
\end{equation}
in the basis $(1A,1B,2A,2B,3A,3B)$.
$h_l$ describes the Dirac electrons in layer $l$,
\begin{equation}
    h_l=\hbar v_F\boldsymbol{\sigma}_l\cdot
    (-\mathrm{i}\nabla-\boldsymbol{\kappa}_l),
    \qquad
    \boldsymbol{\sigma}_l
    =U_l^\dagger(\sigma_x,\sigma_y)U_l,
    \qquad
    U_l=e^{-\mathrm{i}\theta_l\sigma_z/2},
\end{equation}
where $v_F=8.8\times10^5\mathrm{m/s}$ is the Fermi velocity, and $\sigma_{x,y,z}$ are Pauli matrices acting in sublattice space. $\boldsymbol{\kappa}_l=\mathbf{K}_l-\mathbf{K}_2$ denotes the momentum offset of layer $l$ relative to the middle layer, where $\mathbf{K}_l=(2/3,1/3)G_l$ is $K$-point momentum of layer $l$.

The interlayer tunneling operator $T_{l,l+1}$ incorporates the leading momentum dependence arising from the finite in-plane range of interlayer hopping~\cite{SIbib-Yves2025} 
\begin{equation}
    T_{l,l+1}(\mathbf{r})
    =
    \sum_{n=0}^{2}
    W_ne^{\mathrm{i}\mathbf{Q}_{ln}\cdot
    \mathbf{r}}
    \left[
        1+\xi\left(
            \left|\mathbf{K}_2-\mathrm{i}\nabla
            +\mathbf{m}_nG_l\right|
            -|\mathbf{K}_1|
        \right)
    \right],
\end{equation}
where
\begin{equation}
    W_n=
    \begin{pmatrix}
        w_{AA} & w_{AB}e^{-2\pi\mathrm{i}n/3}\\
        w_{AB}e^{2\pi\mathrm{i}n/3} & w_{AA}
    \end{pmatrix},
    \qquad
    \mathbf{Q}_{ln}=\mathbf{m}_n(G_l-G_{l+1}),
\end{equation}
with $\mathbf{m}_0=(0,0)$, $\mathbf{m}_1=(-1,0)$, and $\mathbf{m}_2=(-1,-1)$. We take $w_{AA}=75\,\mathrm{meV}$ and $w_{AB}=110\,\mathrm{meV}$ for the interlayer tunneling amplitudes, and $\xi=-2.1\,\text{\AA}$ for the momentum-dependent correction.
$\mathbf{d}_{12}$ and $\mathbf{d}_{23}$ specify the positions of the two moir\'e patterns relative to a common origin. Their relative displacement, $\mathbf{d}=\mathbf{d}_{23}-\mathbf{d}_{12}$, determines the local domain configuration. Lattice relaxation favors two distinct $C_{3z}$-symmetric configurations, denoted $h$-HTG and $\bar{h}$-HTG, characterized by $\mathbf{d}=\boldsymbol{\delta}$ and $\mathbf{d}=-\boldsymbol{\delta}$, respectively, where
\begin{equation}
    \boldsymbol{\delta}
    =\frac{\mathbf{a}_1+\mathbf{a}_2}{3}, \quad\mathbf{a}_1=\frac{a_0}{{\theta}}(\frac{\sqrt{3}}{2},\frac12),
    \quad\mathbf{a}_2=\frac{a_0}{{\theta}}(0,1).
\end{equation}
Here $\mathbf{a}_{1,2}$ are primitive moir\'e lattice vectors in the small-angle approximation.

For numerical calculations, we construct the continuum Hamiltonian using a circular momentum cutoff of 4.1 times the magnitude of the primitive moiré reciprocal lattice vector. We calculate the top-layer LDOS using a $75\times75$ momentum grid in the moir\'e Brillouin zone. The spin-and valley-summed LDOS is four times the single-spin $K$-valley LDOS, due to the spin degeneracy and time-reversal symmetry. The LDOS spectra are calculated using Gaussian broadening with a standard deviation of $\sigma=2\mathrm{meV}$.

\subsection{Edge-state simulations using a domain-wall structure fitted to the experimental data}
\label{sec:theory_domainwall}
In this section, we investigate the electronic states
near domain walls separating supermoir\'e domains.
Experimentally, domain walls are observed along three directions. We consider a wall approximately parallel to the moir\'e lattice vector $\mathbf a_2$.
To model this domain wall, we introduce a periodic
supercell with lattice vectors
\begin{equation}
    \mathbf L_1=N_{\mathrm s}\mathbf a_1,
    \qquad
    \mathbf L_2=\mathbf a_2,
\end{equation}
where $N_{\mathrm s}$ is the number of moiré unit cells in the supercell.

We extend the single-domain Hamiltonian in
Eq.~\eqref{eq:HTG-Hamiltonian} by allowing the relative
displacement $\mathbf{d}$ to vary across the walls.
The displacement remains approximately constant
within each domain and changes smoothly across
the domain walls.
Using the supercell fractional coordinates defined by
$\mathbf r=u\mathbf L_1+v\mathbf L_2$, we describe this
variation by a double-tanh ansatz independent of $v$:
\begin{equation}
    \mathbf d(\mathbf r)=\mathbf d(u)
    =\mathbf d_{\mathrm i}
    +f(u)\left(\mathbf d_{\mathrm f}-\mathbf d_{\mathrm i}\right),
\end{equation}
where
\begin{equation}
f(u)=\frac{F(u)-F(0)}{F(1/2)-F(0)},
\qquad
F(u)=\frac12
\left[
\tanh\left(w(u-\frac14)\right)
-\tanh\left(w(u-\frac34)\right)
\right].
\end{equation}
Here, $\mathbf d_{\mathrm i}$ and $\mathbf d_{\mathrm f}$ represent the relative displacements of the two domains, and $w$ controls the domain-wall width.

Based on experimental observations, we choose the domain displacements as
\begin{equation}
\mathbf d_{\mathrm i}
=\frac{-\mathbf a_1+2\mathbf a_2}{3}
\quad(\bar h\text{-HTG}),
\qquad
\mathbf d_{\mathrm f}
=\frac{\mathbf a_1+\mathbf a_2}{3}
\quad(h\text{-HTG}),
\end{equation}
and we take
\begin{equation}
    \mathbf d_{12}(\mathbf r)
    =-\frac{\mathbf d(\mathbf r)}{2},
    \qquad
    \mathbf d_{23}(\mathbf r)
    =+\frac{\mathbf d(\mathbf r)}{2}.
\end{equation}
It is worth noting that while only the relative displacement $\mathbf d=\mathbf d_{23}-\mathbf d_{12}$ matters within a single domain, the spatially varying domain-wall structure requires specifying $\mathbf d_{12}(\mathbf r)$ and $\mathbf d_{23}(\mathbf r)$ separately.

\begin{figure}[p]
    \centering
    \includegraphics[width=\textwidth]{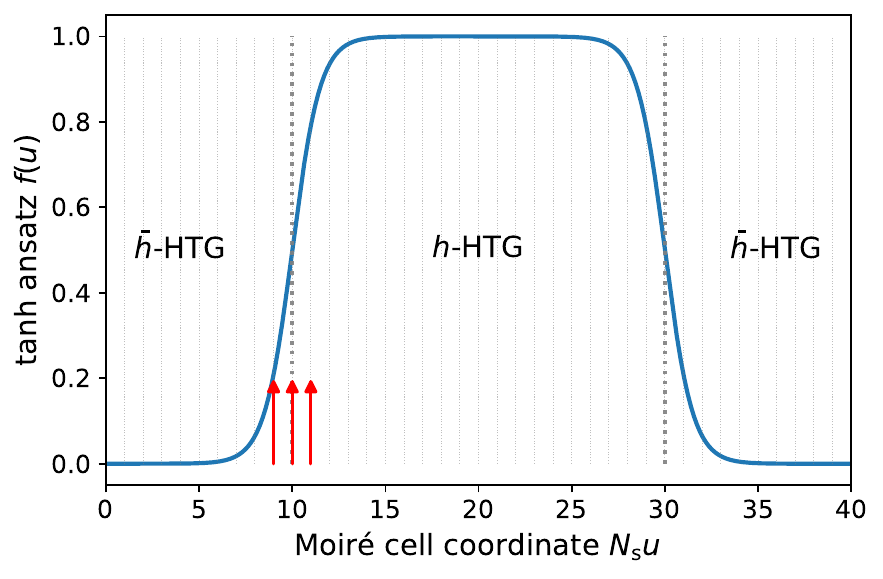}
    \caption{Normalized double-tanh ansatz describing the transition between $\bar h$-HTG and $h$-HTG domains in a periodic supercell with $N_{\mathrm s}=40$ and $w=80/3$. Thin lines mark individual moir\'e cells, while thicker lines indicate the domain-wall centers. Red arrows mark the domain-wall center and the nearest neighboring cell positions used for comparison with experiment.}
    \label{fig:domain_wall}
\end{figure}

We use $N_{\mathrm s}=40$ and $w=80/3$ (\SIfig{fig:domain_wall}).
For numerical calculations, we construct the supercell continuum Hamiltonian using a circular momentum cutoff of 3.1 times the magnitude of the primitive moiré reciprocal lattice vector.
The band dispersion is evaluated along the straight
path from $\mathbf k=\mathbf 0$ to
$\mathbf k=2\pi {\mathbf a}_2/|\mathbf{a}_2|^2$,
parallel to the domain walls at $k_\perp=0$.

We calculate the top-layer LDOS using a $3\times69$
momentum grid in the supercell Brillouin zone and
Gaussian broadening with a standard deviation of
$\sigma=2\,\mathrm{meV}$.
The spectra are evaluated at $(u,v)=(0,0)$ and
$(0.25,0)$, corresponding to the $\bar h$-HTG domain
center and the domain-wall center, respectively.

\section{Moir\'e-site extraction and lateral shifts}
\label{sec:moire-analysis}
\subsection{Large-area supermoir\'e structure}
\label{sec:large-area}

\begin{figure}[p]
  \centering
  \includegraphics[width=\textwidth]{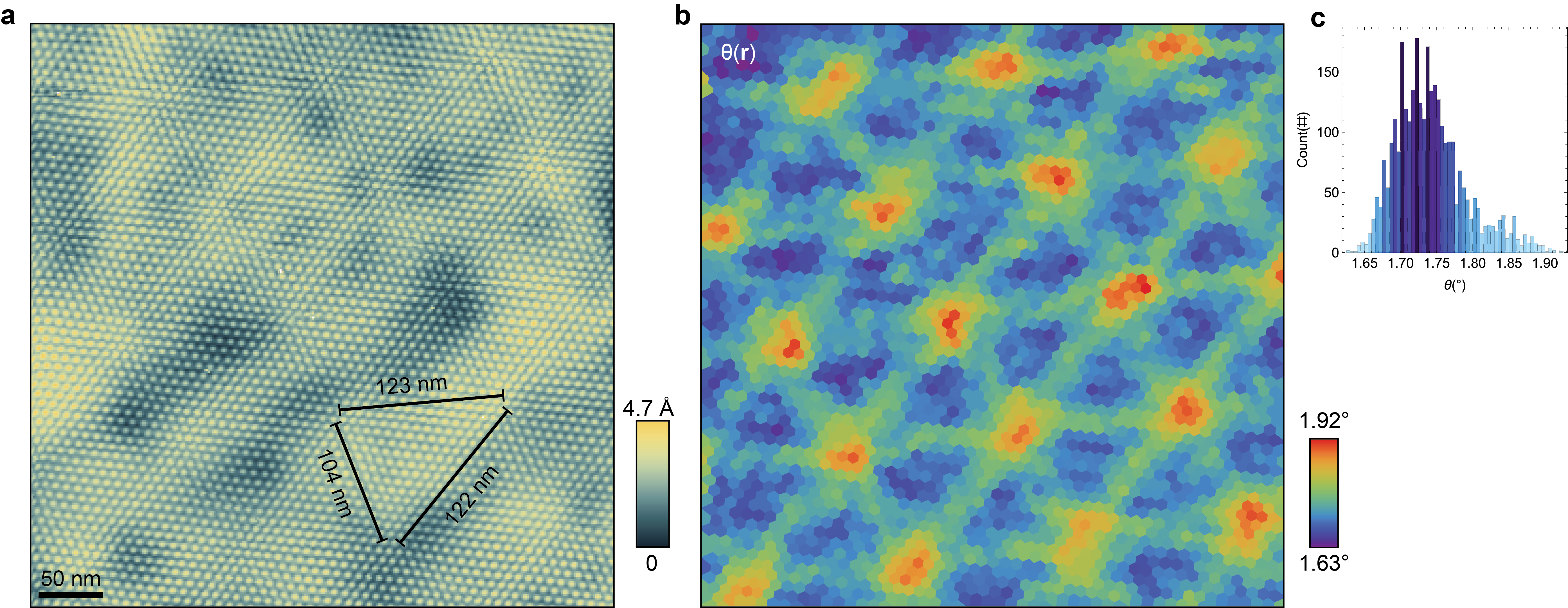}
  \caption{\textbf{Large-area supermoir\'e structure and local moir\'e-derived angle distribution.}
    \textbf{a}, STM topograph resolving the moir\'e modulation and several supermoir\'e domains. The marked triangle has side lengths of 104, 122 and 123~nm. The apparent-height colour scale spans 0--4.7~\AA; scale bar, 50~nm.
    \textbf{b}, Map of the effective local twist angle $\theta(\mathbf r)$ inferred from the moir\'e geometry in the imaged region. The colour scale spans $1.63^\circ$--$1.92^\circ$.
    \textbf{c}, Histogram of the effective angles in \textbf{b}, showing the dominant population near $1.7^\circ$ and a tail towards larger values. The map and histogram characterize local moir\'e geometry rather than separately determining the two interface twist angles. STM acquisition settings: \textbf{a}, sample bias, 500~mV; current setpoint, 10~pA; back-gate voltage, 0~V.}
  \label{fig:large-area}
\end{figure}

The large-area topograph in \SIfig{fig:large-area}a resolves several neighbouring supermoir\'e domains together with the underlying moir\'e modulation. A representative triangular domain has side lengths of 104, 122 and 123~nm, showing that the reconstructed network is not perfectly equilateral. The local moir\'e geometry is expressed as an effective twist-angle map $\theta(\mathbf r)$, obtained by converting the local period extracted from the topograph using $a_m=a/[2\sin(\theta/2)]$, with $a=0.246$~nm (\SIfig{fig:large-area}b). The map shows spatial variations and localized regions with larger effective angles. The corresponding histogram (\SIfig{fig:large-area}c) has its dominant population near $1.7^\circ$, with a tail towards larger angles. These values describe the moir\'e-derived local geometry; in a strained and reconstructed trilayer, they do not independently determine the twist angles of both interfaces. This larger-area characterization complements the domain-scale measurements in \MainFig{fig:one}.

\subsection{Identification of the two moir\'e-site sets}
\label{sec:site-identification}

\begin{figure}[p]
  \centering
  \includegraphics[width=\textwidth]{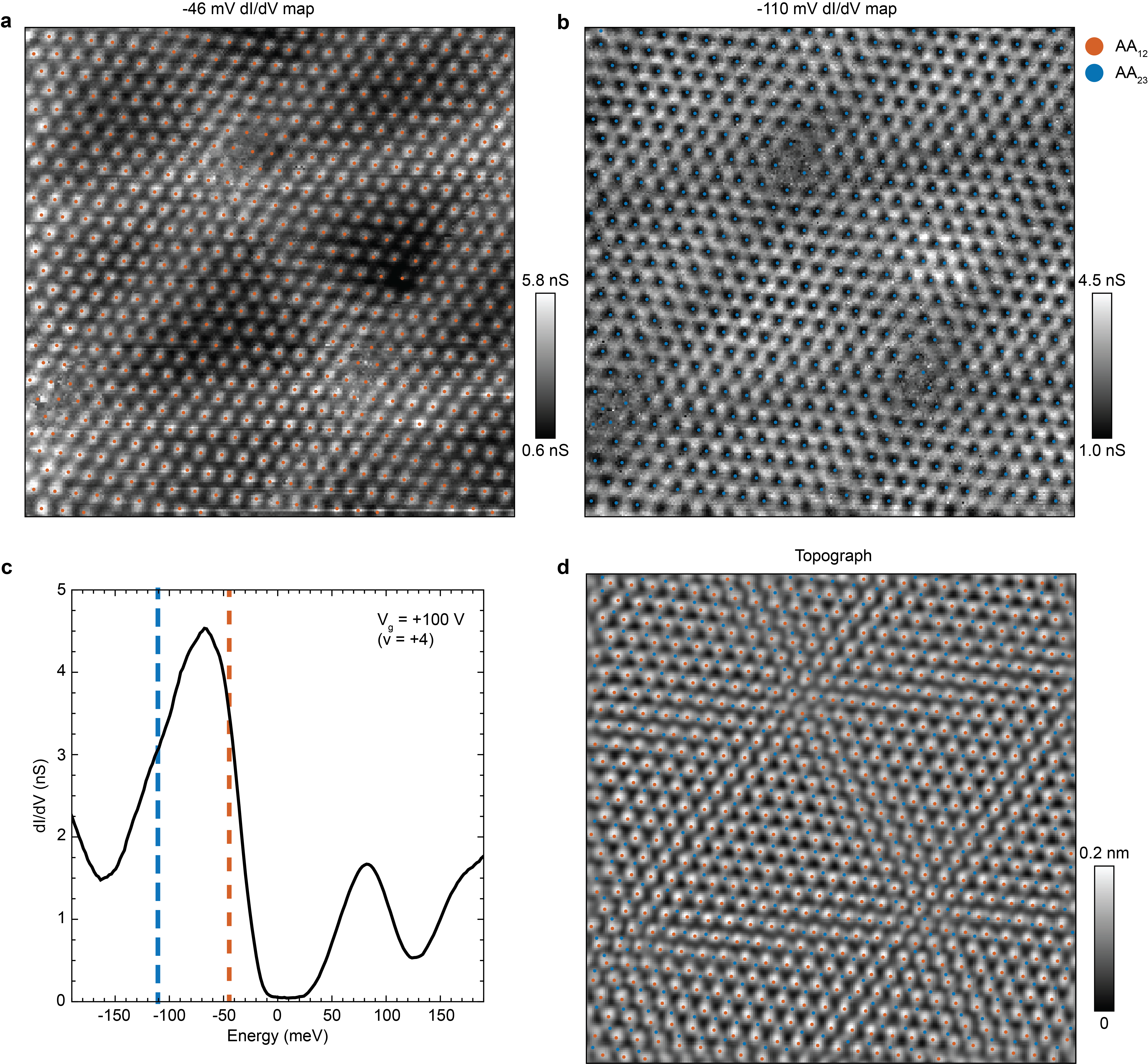}
  \caption{\textbf{Identification and registration of upper- and lower-interface moir\'e centres.}
    \textbf{a}, Differential-conductance map at $V_{\mathrm b}=-46$~mV with the assigned upper-interface \AAt{} centres marked in orange.
    \textbf{b}, Conductance map of the same region at $V_{\mathrm b}=-110$~mV with the assigned lower-interface \AAb{} centres marked in blue.
    \textbf{c}, Representative dI/dV spectrum at $V_{\mathrm g}=+100$~V ($\nu=+4$). Orange and blue dashed lines mark the energies sampled in \textbf{a} and \textbf{b}, respectively.
    \textbf{d}, Corresponding STM topograph with both coordinate sets overlaid, providing the common spatial reference for the relative shifts analysed in \MainFig{fig:one}. STM acquisition settings: \textbf{a-d}, sample bias, 200~mV; current setpoint, 200~pA; back-gate voltage, +100~V; excitation voltage, 2 mV.}
  \label{fig:moire-sites}
\end{figure}

The energy-dependent conductance contrast near the flat bands helps distinguish the upper-interface \AAt{} centres from the lower-interface \AAb{} centres. The surface LDOS differs strongly between these sites, as discussed in \MainFigPanel{fig:two}{c--e}. For example, \SIfig{fig:moire-sites}a,b shows conductance maps of the same region at sample biases of $-46$ and $-110$~mV, respectively, as marked in the dI/dV spectrum (\SIfig{fig:moire-sites}c). The \AAt{} positions are identified as local maxima in the first map and marked in orange, whereas the \AAb{} positions are identified as local minima in the second map and marked in blue. Overlaying both coordinate sets on the corresponding topograph provides a common spatial reference and shows that the extracted positions match the topographic AA-modulation centres (\SIfig{fig:moire-sites}d). The relative arrangement of the two sets changes across the supermoir\'e walls, supplying the coordinates used to construct the lateral shift field in \MainFig{fig:one}.

Spatial averaging gives the following primitive real-space lattice vectors for the upper- and lower-interface moir\'e lattices:

\[\mathbf{a}_{1t} = (8.03651, -0.986805)\text{ nm}; ~\mathbf{a}_{2t} = (5.02248, 6.47195)\text{ nm}\]
\[\mathbf{a}_{1b} = (8.59009, -1.22576)\text{ nm}; ~\mathbf{a}_{2b} = (5.37308, 6.92728)\text{ nm}\]

where the subscripts t and b denote the top--middle and middle--bottom interfaces, respectively.

\section{Extraction of the strain configuration}
\label{sec:strain_config}

We infer the layer-strain configurations from the measured moir\'e lattice vectors and top-layer atomic orientation, assuming uniaxial strain with a Poisson response in each graphene sheet. The labels top, middle and bottom refer to the three sheets, denoted by t, m and b, respectively. The column-vector convention used below differs from the row-vector convention used in Supplementary Section~\ref{sec:theory_singledomain}.

\subsection{Input geometry and interpretation}

All vectors are expressed in one Cartesian coordinate frame, with angles
measured counterclockwise from \(+x\). Primitive vectors form the
columns of each basis matrix. The supplied vectors are
real-space lattice vectors, rather than reciprocal wavevectors.
Using the interface assignments established in Supplementary Section~\ref{sec:site-identification},
\begin{align}
 A_{\rm t}
 &= [\,\mathbf a_{1t}\ \mathbf a_{2t}\,]
 =\begin{pmatrix}
 8.03651 & 5.02248\\[-1pt]
 -0.986805 & 6.47195
 \end{pmatrix}\nm ,
 &&\text{top--middle \moire}, \label{eq:At}\\
 A_{\rm b}
 &= [\,\mathbf a_{1b}\ \mathbf a_{2b}\,]
 =\begin{pmatrix}
 8.59009 & 5.37308\\[-1pt]
 -1.22576 & 6.92728
 \end{pmatrix}\nm ,
 &&\text{middle--bottom \moire}. \label{eq:Ab}
\end{align}
The top-sheet atomic vectors measured from \MainFigPanel{fig:two}{b} are
\begin{equation}
 \mathbf l_1=
 \begin{pmatrix}0.224005\\0.101676\end{pmatrix}\nm,
 \qquad
 \mathbf l_2=
 \begin{pmatrix}0.0238303\\0.251012\end{pmatrix}\nm .
 \label{eq:atomicinputs}
\end{equation}
Their lengths and measured cell distortion are excluded from the fit,
as these may include scan drift. Only the direction of \(\mathbf l_1\)
is imposed as the additional atomic constraint: $\phi_1=\atan(l_{1y},l_{1x})$. The direction of \(\mathbf l_2\) is reserved for a separate comparison,
and is not imposed simultaneously.

The reference graphene lattice is the regular hexagonal Bravais lattice
with two primitive vectors separated by \(60^\circ\):
\begin{equation}
 A_{0}^{\rm G}
 =a\begin{pmatrix}
 1&1/2\\
 0&\sqrt{3}/2
 \end{pmatrix},
 \qquad a=0.246\,\nm .
 \label{eq:reference}
\end{equation}
Regularity is imposed on this unit cell. Physical strain
in all three sheets remains an unknown; the distorted measured atomic
cell in Eq.~\eqref{eq:atomicinputs} is not interpreted as physical strain.

\subsection{Uniaxial deformation model}

The two-dimensional rotation matrix is
\begin{equation}
 R(\gamma)=
 \begin{pmatrix}\cos\gamma&-\sin\gamma\\
                 \sin\gamma& \cos\gamma\end{pmatrix}.
 \label{eq:rotation}
\end{equation}
The stretch of a sheet \(\ell\in\{t,m,b\}\) is
\begin{equation}
 S_\ell
 =R(\alpha_\ell)
 \begin{pmatrix}
 1+\epsilon_\ell&0\\
 0&1-\nu_{\rm P}\epsilon_\ell
 \end{pmatrix}
 R(\alpha_\ell)^T,
 \qquad \nu_{\rm P}=0.16.
 \label{eq:stretch}
\end{equation}
Here, $\nu_{\rm P}$ is the Poisson ratio, distinct from the electronic filling $\nu$. The uniaxial strain $\epsilon_\ell$ is positive for tension and negative for compression.
This deformation and the reciprocal-lattice difference construction
are standard in HTG strain modelling~\cite{SIbib-hoke2026relaxation}.

The full atomic basis becomes
\begin{equation}
 A^{\rm G}_\ell=S_\ell R(\beta_\ell)A^{\rm G}_{0}.
 \label{eq:atomicbasis}
\end{equation}
The order in Eq.~\eqref{eq:atomicbasis} fixes the convention:
\(\beta_\ell\) is the rigid rotation angle of the graphene sheet, and
\(S_\ell\) acts along laboratory-frame axes. No independent isotropic strain term is introduced.

\subsection{Reciprocal-lattice reconstruction}

For any graphene-lattice basis \(A^{\rm G}\), its reciprocal basis is
\begin{equation}
 G=2\pi (A^{\rm G})^{-T},
 \qquad (A^{\rm G})^T G=2\pi\Id .
 \label{eq:reciprocal}
\end{equation}
Consequently, the measured \moire\ reciprocal bases are
\begin{equation}
 Q_{\rm t}=2\pi A_{\rm t}^{-T},
 \qquad Q_{\rm b}=2\pi A_{\rm b}^{-T}.
 \label{eq:moireQ}
\end{equation}

The measured moir\'e bases and the graphene reciprocal bases use different primitive-vector orderings. With the present convention, they are related by
\begin{equation}
 G_{\rm m}-G_{\rm t}=Q_{\rm t}M,
 \qquad
 G_{\rm b}-G_{\rm m}=Q_{\rm b}M,
 \qquad
 M=\begin{pmatrix}0&1\\-1&1\end{pmatrix}.
 \label{eq:basis-correspondence}
\end{equation}
Thus, once the top-sheet reciprocal basis is specified, the middle- and bottom-sheet bases follow successively from these relations. The real-space bases are recovered using $A^{\rm G}_\ell=2\pi G_\ell^{-T}$.

\subsection{Using the atomic direction without its measured distortion}

The measured angle $\phi_1$ of $\mathbf{l}_1$ is the direction of
the first strained primitive vector. It reflects both strain and the rigid sheet rotation \(\beta_t\).
Applying the inverse strain operator to this direction gives the unit vector $\mathbf v_t$ of the rigidly rotated sheet:
\begin{equation}
 \mathbf u_1=\begin{pmatrix}\cos\phi_1\\\sin\phi_1\end{pmatrix},
 \qquad
 \mathbf v_t=
 \frac{S_t^{-1}\mathbf u_1}
      {|S_t^{-1}\mathbf u_1|}.
 \label{eq:orientationinverse}
\end{equation}
The top-sheet rigid rotation is therefore fixed by $\beta_t=\atan(v_{t,y},v_{t,x})$.

\subsection{Imposing uniaxial strain on the other sheets}

For each reconstructed sheet, we obtain the deformation and its left polar decomposition as
\begin{equation}
 F_\ell=A^{\rm G}_\ell(A^{\rm G}_{0})^{-1},
 \qquad
 S_\ell=(F_\ell F_\ell^T)^{1/2},
 \qquad
 R_\ell=S_\ell^{-1}F_\ell .
 \label{eq:polar}
\end{equation}

The principal stretches are constrained to be $1+\epsilon_\ell$ and $1-\nu_{\rm P}\epsilon_\ell$. The eigenvector associated with $1+\epsilon_\ell$ defines the strain direction $\alpha_\ell$. The rigid sheet rotation angle is
\begin{equation}
 \beta_\ell=\atan\bigl((R_\ell)_{21},(R_\ell)_{11}\bigr).
 \label{eq:rotationextract}
\end{equation}

\subsection{Strain configurations and interface twists}

Using this procedure, we obtain two isolated strain configurations that satisfy the measured \moire\
cells and the \(\mathbf l_1\) direction constraint.
Table~\ref{tab:solutions} lists the parameters in
Eq.~\eqref{eq:atomicbasis}. Strains are displayed as percentages. All strain-axis and sheet-rotation angles are in degrees.

\begin{table}[htbp]
 \centering
 \small
 \setlength{\tabcolsep}{5pt}
 \begin{tabular}{lrrrrrr}
  \toprule
  &\multicolumn{3}{c}{Solution 1}&\multicolumn{3}{c}{Solution 2}\\
  Layer
  &\(\epsilon_\ell(\%)\)&\(\alpha_\ell\)&\(\beta_\ell\)
       &\(\epsilon_\ell(\%)\)&\(\alpha_\ell\)&\(\beta_\ell\)\\
  \midrule
  Top    & \(-0.049222\)&151.126902&24.397644
         & \(-0.101724\)&89.214381&24.439393\\
  Middle & \(+0.093550\)&74.731167&22.659827
         & \(+0.046156\)&179.476802&22.701966\\
  Bottom & \(+0.091390\)&84.130333&21.045687
         & \(+0.048954\)&158.547258&21.088130\\
  \bottomrule
 \end{tabular}
 \caption{Two configurations with uniaxial strain in each graphene sheet. The longitudinal strain is $\epsilon_\ell$, the strain-axis angle relative to $+x$ is $\alpha_\ell$, and the rigid sheet rotation relative to $+x$ is $\beta_\ell$.
 Negative longitudinal strain denotes compression.}
 \label{tab:solutions}
\end{table}

The signed interface twists are the differences between the rigid sheet rotations $\beta_\ell$:
\begin{equation}
 \theta_{\rm t}=\beta_m-\beta_t,
 \qquad
 \theta_{\rm b}=\beta_b-\beta_m.
 \label{eq:twists}
\end{equation}
Both are negative in the chosen coordinate frame, corresponding to
clockwise rotation from top to middle to bottom.
The twist-magnitude mismatch is
\begin{equation}
 \Delta\theta=|\theta_{\rm t}|-|\theta_{\rm b}|.
 \label{eq:twistdifference}
\end{equation}

\begin{table}[htbp]
 \centering
 \begin{tabular}{lrr}
  \toprule
  Quantity&Solution 1&Solution 2\\
  \midrule
  \(\theta_{\rm t}\)&\(-1.737817^\circ\)&\(-1.737427^\circ\)\\
  \(\theta_{\rm b}\)&\(-1.614140^\circ\)&\(-1.613836^\circ\)\\
  \(\Delta\theta\)&\(0.123677^\circ\)&\(0.123591^\circ\)\\
  \bottomrule
 \end{tabular}
 \caption{Interface twists extracted from the rigid polar rotations.}
 \label{tab:twists}
\end{table}

\subsection{Comparison with the second atomic direction to find unique solution}

The measured direction of the second atomic vector is
\begin{equation}
 \phi_2^{\rm meas}=\atan(l_{2y},l_{2x})
                 =84.5767708081^\circ .
 \label{eq:phi2measured}
\end{equation}
For each reconstructed top sheet, its predicted second direction
and angular mismatch are
\begin{equation}
 \phi_2^{\rm pred}=\arg(A^{\rm G}_t\mathbf e_2),
 \qquad
 \delta\phi_2=
 \left|
       \phi_2^{\rm pred}-\phi_2^{\rm meas}\right|,
 \quad\mathbf e_2=(0,1)^T .
 \label{eq:phi2check}
\end{equation}
The comparison gives

\begin{table}[htbp]
 \centering
 \begin{tabular}{lrr}
  \toprule
  &\(\phi_2^{\rm pred}\)&\(\delta\phi_2\)\\
  \midrule
  Solution 1&\(84.385770^\circ\)&\(0.191001^\circ\)\\
  Solution 2&\(84.433778^\circ\)&\(0.142992^\circ\)\\
  \bottomrule
 \end{tabular}
 \caption{Comparison with the measured \(\mathbf l_2\) direction, which
 was not used to obtain the strain solutions.}
 \label{tab:l2check}
\end{table}

Solution 2 gives closer agreement with the second atomic direction, by $0.048008^\circ$, and is therefore the preferred solution quoted in the main text.

\section{Gate-dependent spectroscopy and noise assessment}
\label{sec:spectroscopy}
\begin{figure}[p]
  \centering
  
  \begin{overpic}[width=\textwidth]{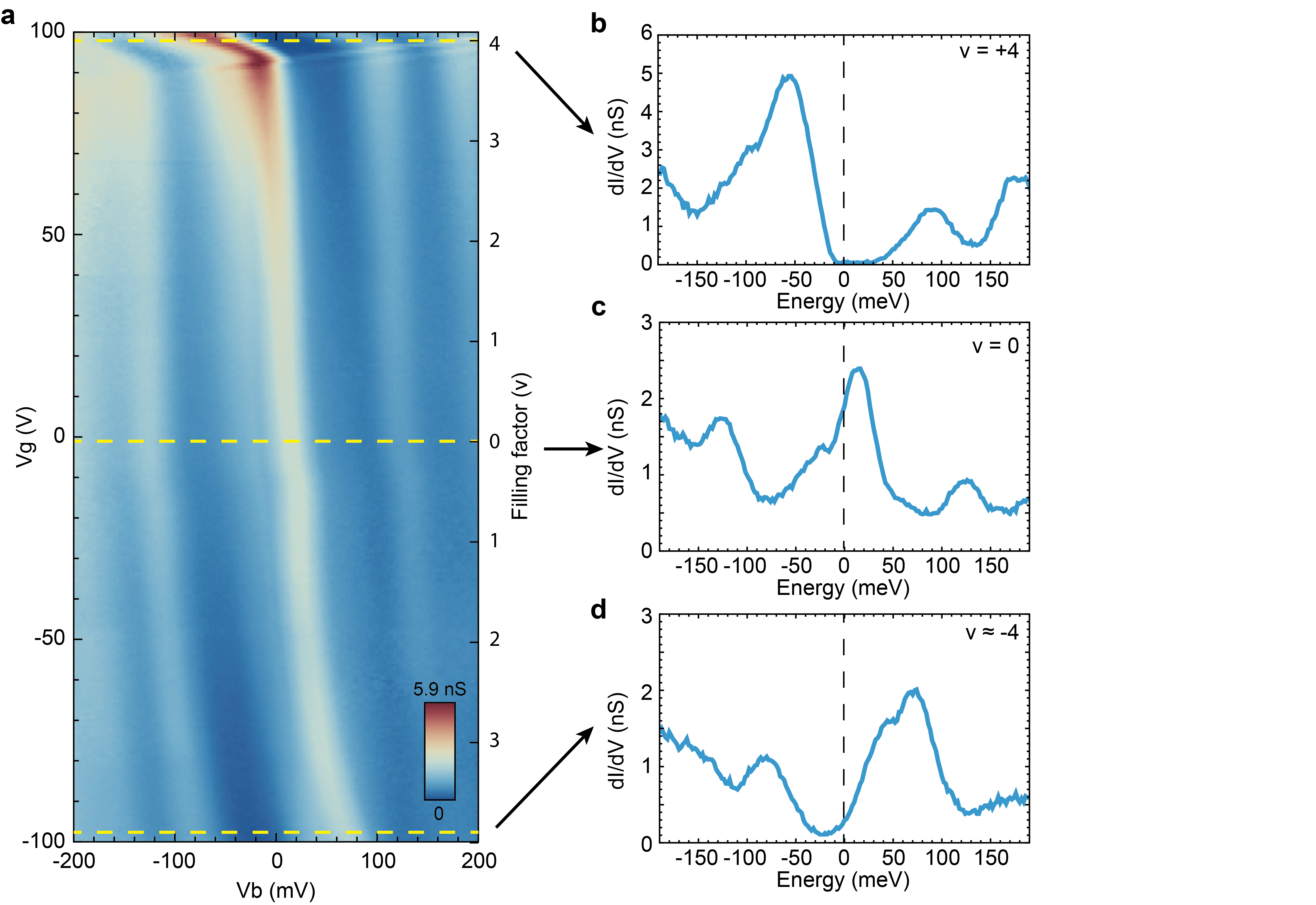}
    \put(36.7,26){\makebox(0,0)[bl]{\color{white}\rule{0.020\textwidth}{0.038\textwidth}}}
    \put(37.7,27.8){\makebox(0,0){\fontfamily{phv}\fontsize{6.2}{7}\selectfont\textminus1}}
    \put(36.7,18.4){\makebox(0,0)[bl]{\color{white}\rule{0.020\textwidth}{0.038\textwidth}}}
    \put(37.7,20.2){\makebox(0,0){\fontfamily{phv}\fontsize{6.2}{7}\selectfont\textminus2}}
    \put(36.7,10.8){\makebox(0,0)[bl]{\color{white}\rule{0.020\textwidth}{0.038\textwidth}}}
    \put(37.7,12.6){\makebox(0,0){\fontfamily{phv}\fontsize{6.2}{7}\selectfont\textminus3}}
  \end{overpic}
  \caption{\textbf{Gate tuning of the flat-band resonances and remote-band isolation gaps.}
    \textbf{a}, Differential-conductance map versus sample bias $V_{\mathrm b}$ and gate voltage $V_{\mathrm g}$, spanning $-200$ to $+200$~mV and approximately $-100$ to $+100$~V, respectively. Colour represents differential conductance. Yellow dashed lines indicate the three gate-voltage cuts shown in \textbf{b--d}, near $+100$, 0 and $-100$~V.
    \textbf{b--d}, Spectra at the marked cuts, labelled $\nu=+4$ (\textbf{b}), $\nu=0$ (\textbf{c}) and $\nu\simeq-4$ (\textbf{d}). The vertical dashed line in each spectrum marks the Fermi level, $E=0$. The electron-side isolation gap is aligned with the Fermi level at $\nu=+4$, while the hole-side conductance minimum approaches it near $\nu\simeq-4$. Differential conductance is plotted in nS. STM acquisition settings: sample bias, 200~mV; current setpoint, 200~pA; excitation voltage, 2 mV.}
  \label{fig:gate-spectroscopy}
\end{figure}

The gate-dependent conductance map in \SIfig{fig:gate-spectroscopy}a tracks the evolution of the flat-band and remote-band features as $V_{\mathrm g}$ is swept from approximately $-100$~V to $+100$~V. With increasing gate voltage, the central flat-band feature shifts from positive to negative sample bias. Spectra along the three marked gate-voltage cuts are shown in \SIfig{fig:gate-spectroscopy}b--d. At $\nu=0$, the flat-band feature lies close to the Fermi level. At $\nu=+4$, the electron-side isolation gap is brought to the Fermi level and the differential conductance is strongly suppressed over a finite energy interval. Near $\nu\simeq-4$, the Fermi level lies in the corresponding hole-side conductance minimum, which retains a larger residual background. These spectra establish the gate settings used to access the remote-band gaps for the domain-wall comparisons in \MainFigPanel{fig:three}{d,e}.

\begin{figure}[p]
  \centering
  \includegraphics[width=\textwidth]{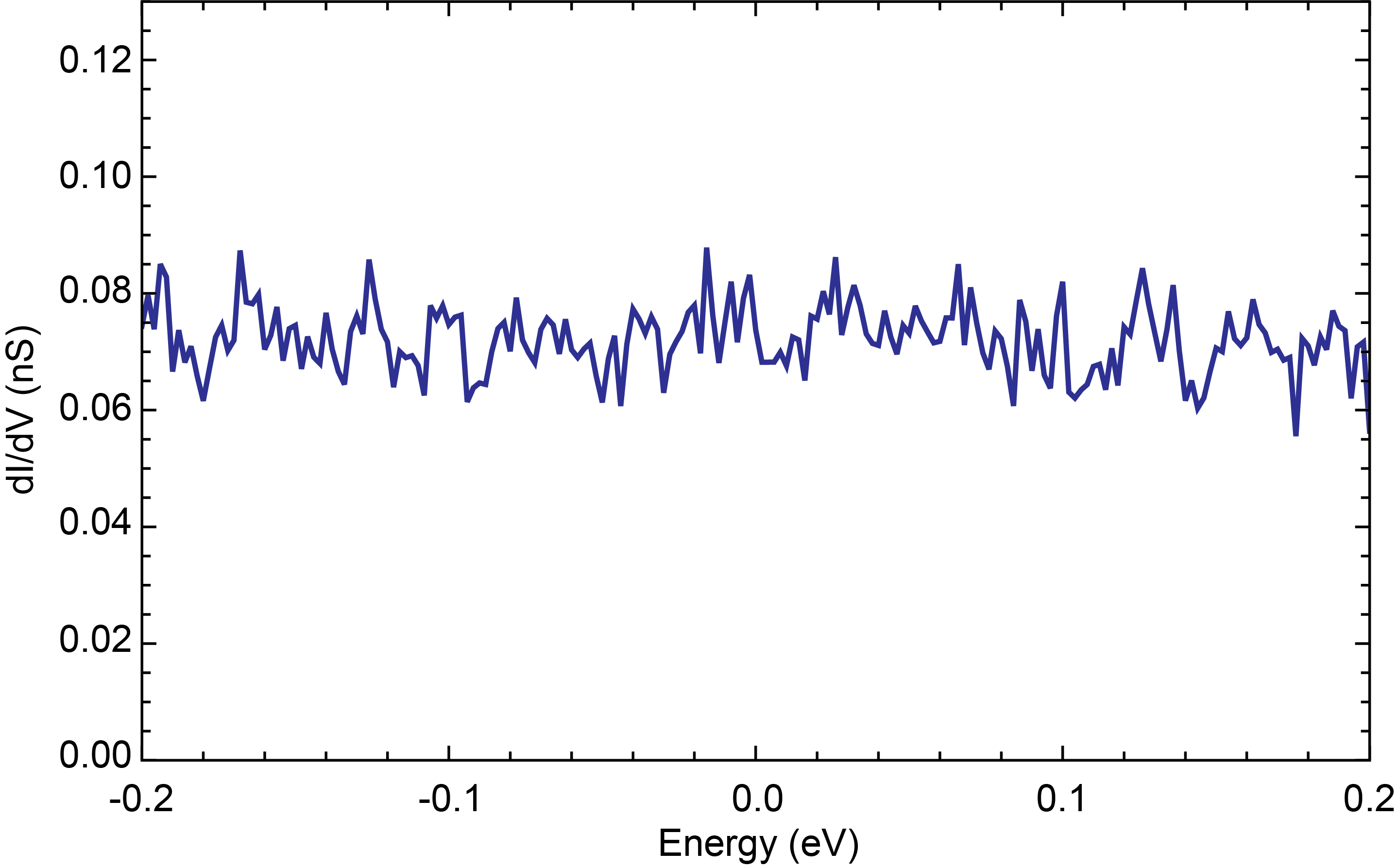}
  \caption{\textbf{Retracted-tip background and instrumental noise in dI/dV spectroscopy.}
    Differential conductance measured with the tip retracted from the sample. The approximately bias-independent mean background is 0.07~nS; fluctuations about this level characterize the instrumental noise. The nominal spectroscopy settings before tip retraction were a sample bias of 0.2~V and a current setpoint of 200~pA; the excitation voltage was 2~mV. The displayed background trace is shown before background subtraction. STM acquisition settings: sample bias, 200~mV; current setpoint, 200~pA; excitation voltage, 2 mV.}
  \label{fig:noise}
\end{figure}

We characterize the dI/dV background by recording a spectrum with the tip retracted from the sample, using the same acquisition settings as for the HTG spectroscopy. An example is shown in \SIfig{fig:noise}, with a mean background of approximately 0.07~nS. The experimental HTG dI/dV spectra presented in this work have been corrected by subtracting this background level. The residual fluctuations in the retracted-tip spectrum provide a measure of the instrumental noise.

\section{Electronic states at a substrate step}

\begin{figure}[p]
\centering
\includegraphics[width=\textwidth]{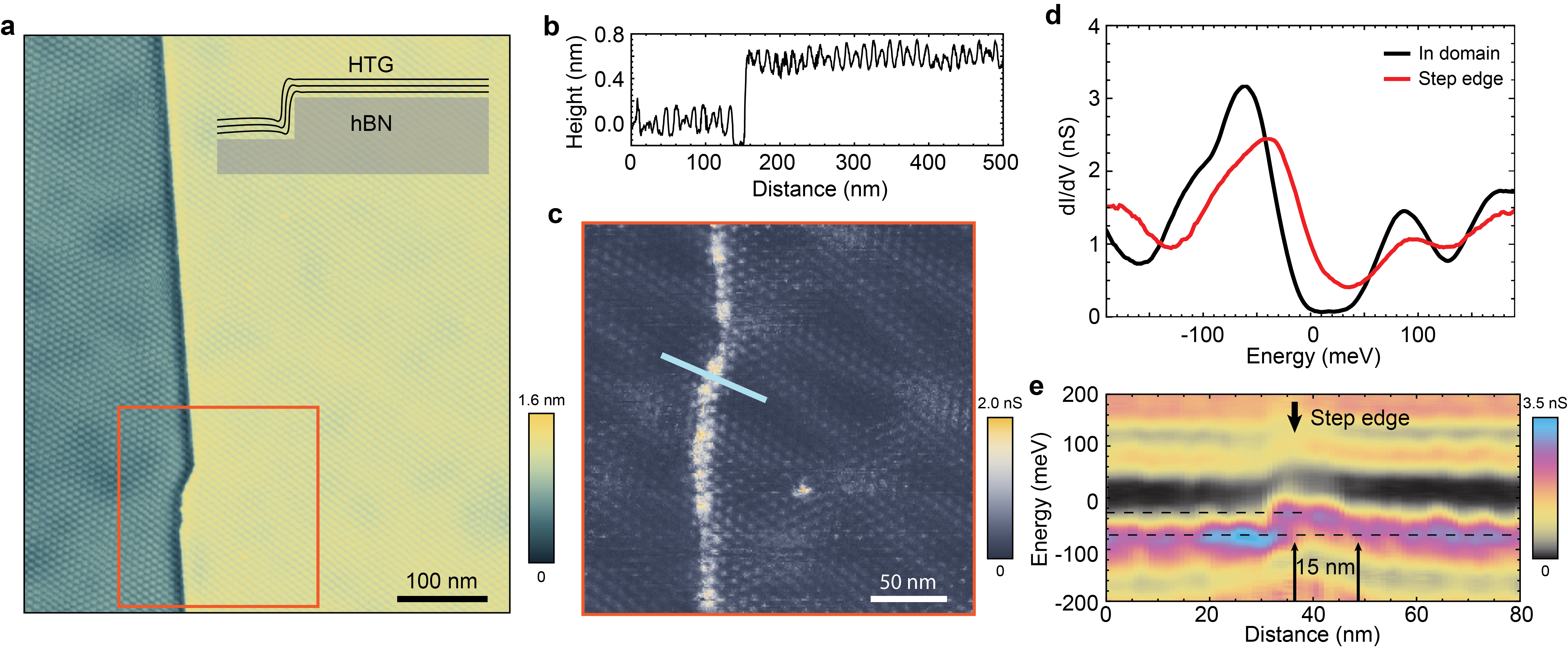}
\caption{\textbf{In-gap states at an hBN substrate step.}
\textbf{a}, STM topograph of HTG crossing a substrate step. Inset, schematic of a continuous trilayer draped over the hBN step. The orange rectangle identifies the region shown in \textbf{c}. \textbf{b}, Topographic-height profile across the step, with a change of approximately 0.6~nm, consistent with two hBN layers. \textbf{c}, dI/dV map at an energy inside the domain-interior remote-band gap ($\nu=+4$), showing enhanced LDOS weight along the step. The light-blue line marks the spectroscopy path in \textbf{e}.
\textbf{d}, Spectra in a domain interior (black) and at the step (red), showing increased in-gap LDOS and a shifted flat-band feature at the step edge.
\textbf{e}, dI/dV versus energy and position along the line in \textbf{c}. The downward arrow marks the position of the step edge; dashed horizontal lines provide energy references, and the two upward arrows delimit the indicated 15~nm spatial interval of the spectral change. STM acquisition settings: \textbf{a,b}, sample bias, 200~mV; current setpoint, 100~pA; \textbf{c-e}, sample bias, 200~mV; current setpoint, 200~pA; back gate voltage: +100~V, excitation voltage: 2 mV.} 
\label{fig:step_edge}
\end{figure}

An hBN substrate step provides a distinct boundary geometry (\SIfig{fig:step_edge}a). The topographic height profile changes by approximately $0.6$~nm (\SIfig{fig:step_edge}b), consistent with a two-layer hBN step. The trilayer appears to continue across it, as illustrated in the inset of \SIfig{fig:step_edge}a. The step therefore defines an internal interface rather than an exposed graphene termination. Deformation and electrostatic changes at substrate steps can produce a potential barrier in graphene\cite{SIbib-Low2012}, motivating an examination of the local remote-band-gap spectrum at this interface.

The in-gap conductance is strongly enhanced along the step (\SIfig{fig:step_edge}c). Compared with a domain-interior spectrum, the step spectrum has greater low-energy weight and a shifted flat-band feature (\SIfig{fig:step_edge}d). The spectra and energy--position map show a local upward shift of the flat-band feature by approximately 30~meV, accompanied by increased weight in the surrounding low-conductance interval (\SIfig{fig:step_edge}e). The spatial change occurs over roughly 15--20~nm, comparable to two moir\'e periods. These observations demonstrate a localized electronic reconstruction at the substrate step.

The in-gap response at the substrate step is stronger and more localized than at the supermoir\'e domain walls visible in the same map (\SIfig{fig:step_edge}c). This difference may reflect a sharper local potential at the substrate step. Studies of other two-dimensional topological interfaces also show that confinement and electrostatics can reshape boundary spectra\cite{SIbib-Van2020,SIbib-Jiachen2025}. In the present geometry, the observed band shift is consistent with a step-induced potential\cite{SIbib-Low2012}. Because the graphene continues across the step, these measurements do not by themselves establish a topological origin for the step-associated in-gap states.

\clearpage

\clearpage

\clearpage

\clearpage

\clearpage

\clearpage

\end{document}